\documentclass{SciPost}

\hypersetup{
    colorlinks,
    linkcolor={red!50!black},
    citecolor={blue!50!black},
    urlcolor={blue!80!black}
}

\usepackage[bitstream-charter]{mathdesign}
\DeclareSymbolFont{usualmathcal}{OMS}{cmsy}{m}{n}
\DeclareSymbolFontAlphabet{\mathcal}{usualmathcal}

\fancypagestyle{SPstyle}{
\fancyhf{}
\lhead{\colorbox{scipostblue}{\bf \color{white} ~SciPost Physics }}
\rhead{{\bf \color{scipostdeepblue} ~Submission }}

\fancyfoot[C]{\textbf{\thepage}}
}

\usepackage{bm}

\newcommand{\bra}[1]{\ensuremath{\langle#1|}}
\newcommand{\ket}[1]{\ensuremath{|#1\rangle}}

\newcommand{\tr}{{\rm Tr}}

\newcommand{\vect}[1]{\bm{#1}}

\newcommand{\overbar}[1]{\mkern 1.5mu\overline{\mkern-1.5mu#1\mkern-1.5mu}\mkern 1.5mu}
\newcommand{\n}{\nonumber\\}
\newcommand{\ex}[1]{\langle #1\rangle}

\newcommand{\be}{\begin{equation}}
\newcommand{\ee}{\end{equation}}

\newcommand{\beq}{\begin{eqnarray}}
\newcommand{\eeq}{\end{eqnarray}}

\allowdisplaybreaks[3]

\begin{document}

\pagestyle{SPstyle}

\begin{center}{\Large \textbf{\color{scipostdeepblue}{
%%%%%%%%%% TODO: Write your article's title here
Quantum Chaos, Randomness and Universal Scaling of Entanglement in Various Krylov Spaces
}}}\end{center}

\begin{center}\textbf{
% Use (full) first name (+ middle name initials) + surname format.
% Separate subsequent authors by a comma, omit comma and use "and" for the last author.
% Mark the corresponding author(s) with a superscript symbol in this order
% \star, \dagger, \ddagger, \circ, \S, \P, \parallel, ...
Hai-Long Shi\textsuperscript{1$\star$},
Augusto Smerzi\textsuperscript{1$\dagger$} and
Luca Pezz\`e\textsuperscript{1$\ddagger$}
}\end{center}

\begin{center}
{\bf 1} QSTAR, INO-CNR and LENS, Largo Enrico Fermi 2, 50125 Firenze, Italy
%%%%%%%%%% END TODO: AFFILIATIONS
% Provide email address of corresponding author(s)
\\[\baselineskip]
$\star$ \href{mailto:hailong.shi@ino.cnr.it}{\small hailong.shi@ino.cnr.it}\\
$\dagger$ \href{mailto:augusto.smerzi@ino.cnr.it}{\small augusto.smerzi@ino.cnr.it}\,,\quad
$\ddagger$ \href{mailto:luca.pezze@ino.cnr.it}{\small luca.pezze@ino.cnr.it}
\end{center}

\section*{\color{scipostdeepblue}{Abstract}}
\textbf{\boldmath{%
Multipartite entanglement is a crucial resource for advancing quantum technologies, with considerable research efforts directed toward achieving its rapid and scalable generation.
In this work, we derive an analytical expression for the time-averaged quantum Fisher information (QFI), enabling the detection of scalable multipartite entanglement dynamically generated by all quantum chaotic systems governed by Dyson's ensembles.
Our approach integrates concepts of randomness and quantum chaos, 
demonstrating that the QFI is universally determined by the structure and dimension of the Krylov space that confines the chaotic dynamics.
In particular, the QFI ranges from $N^2/3$ for $N$ qubits in the permutation-symmetric subspace (e.g. for  chaotic kicked top models with long-range interactions), to $N$ when the dynamics extend over the full Hilbert space with or without bit reversal
symmetry or parity symmetry (e.g. in chaotic models with short-range Ising-like interactions).
In the former case, the QFI reveals multipartite entanglement among $N/3$ qubits and highlights the power of chaotic collective spin systems in generating scalable multipartite entanglement. 
Interestingly this result can be related to isotropic substructures in the Wigner distribution of chaotic states and demonstrates the efficacy of quantum chaos for Heisenberg-scaling quantum metrology.
Finally, our general expression for the QFI agrees with that obtained for random states and, differently from out-of-time-order-correlators, it can also distinguish chaotic from integrable unstable
spin dynamics. 
}}

\vspace{\baselineskip}

%%%%%%%%%% BLOCK: Copyright information
% This block will be filled during the proof stage, and finilized just before publication.
% It exists here only as a placeholder, and should not be modified by authors.
\noindent\textcolor{white!90!black}{%
\fbox{\parbox{0.975\linewidth}{%
\textcolor{white!40!black}{\begin{tabular}{lr}%
  \begin{minipage}{0.6\textwidth}%
    {\small Copyright attribution to authors. \newline
    This work is a submission to SciPost Physics. \newline
    License information to appear upon publication. \newline
    Publication information to appear upon publication.}
  \end{minipage} & \begin{minipage}{0.4\textwidth}
    {\small Received Date \newline Accepted Date \newline Published Date}%
  \end{minipage}
\end{tabular}}
}}
}
%%%%%%%%%% BLOCK: Copyright information

% For convenience during refereeing we turn on line numbers:
%\linenumbers

\vspace{10pt}
\noindent\rule{\textwidth}{1pt}
\tableofcontents
\noindent\rule{\textwidth}{1pt}
\vspace{10pt}

% Write your article contents here, starting from first \section.
% An example structure is given below.

\section{Introduction}
\label{sec:intro}
% TODO: write your article here.
Engineering large multipartite entanglement in ensembles of many qubits is central to quantum simulations~\cite{GeorgescuRMP2014}, information theory~\cite{HorodeckiRMP2009}, and metrology~\cite{PezzeRMP2018}.
These states are also pivotal for enhancing our understanding of condensed matter~\cite{AmicoRMP2001, LaflorenciePhysRep2016} and high-energy physics~\cite{WittenRMP2018,RyuJHEP2006}.
One of the primary objectives is the rapid generation of multipartite entanglement from readily available non-entangled states, such as (Gaussian) coherent spin states~\cite{ArecchiPRA1972}, using many-body dynamics~\cite{KitagawaPRA1993, DefenuPR2024,MicheliPRA2003, MuesselPRA2015,SorelliPRA2019,HuangPRA2022}.

The quantum Fisher information (QFI) not only quantifies the ultimate quantum sensing ability but also serves as a powerful detector of multipartite entanglement~\cite{PezzePRL2009,HyllusPRA2012,TothPRA2012} in non-Gaussian states~\cite{StrobelSCIENCE2014,BohnetSCIENCE2016}.
The established connection~\cite{GarttnerPRL2018, PappalardiPRA2018, LerosePRA2020} between out-of-time-order correlators (OTOCs)—a measure of information scrambling—and the QFI expands the scope of chaotic quantum systems~\cite{WimbergerBook2014}  expected to exhibit exponentially fast generation of multipartite entanglement, owing to their ability to scramble information at an exponential rate.
However, due to another constraint imposed by the Lieb-Robinson bounds on operator dynamics~\cite{TrainPRL2021,GuoPRA2020}, this does not generally imply that scrambling dynamics leads to scalable multipartite entanglement~\cite{ShiPRL2024,ChuPRL2023,ChuArXiv2024} in the long run.
It thus remains unclear which types of chaotic systems can generate scalable multipartite entanglement, i.e., QFI scaling as $N^2$  known as the Heisenberg scaling.

In this Letter, we investigate multipartite entanglement generated by chaotic quantum dynamics for times longer than the scrambling (or Ehrenfest) time $t^*$. 
In particular, we study the time-averaged QFI,
\begin{flalign} \label{Fchaos}
\bar{F}_{\rm chaos}[\hat O]\equiv \lim_{T\to\infty} \frac{1}{T-t^*}\int_{t^*}^{T} dt F_Q[\ket{\psi_{\rm chaos}(t)}, \hat O],
\end{flalign}
where $F_Q[\ket{\psi_{\rm chaos}(t)},\hat O]$ is the QFI of the quantum state $\ket{\psi_{\rm chaos}(t)}$ evolved under chaotic dynamics and $\hat O$ is a generic Hermitian operator. 
We compute analytically Eq.~(\ref{Fchaos}) based on standard random matrix theory (RMT) for all chaotic models described by Dyson's three circular ensembles~\cite{DysonMathPhys1962-1,DysonMathPhys1962-2,DysonMathPhys1962-3} and using the ergodicity hypothesis~\cite{GoldsteinEPJH2010,PolkovnikovRMP2011}.
We demonstrate the universal behavior of QFI in chaotic systems
\begin{flalign}\label{Main}
\bar{F}_{\rm chaos}[\hat{O}]\!=\! 
\frac{4\tr_{\mathcal K}[\hat{O}^2]}{K}
\!-\!
\frac{4 \tr_{\mathcal K}[\hat O]^2}{K^2}
\!+\!
\mathcal O\bigg(\frac{\tr_{\mathcal K}[\hat{O}^2]}{K^2}\bigg),
\end{flalign}
which is universally governed by the structure and dimension $K\!\equiv\!{\rm dim}\mathcal K $ of the Krylov space $\mathcal K$.
The Krylov space is the minimal subspace to confine the chaotic dynamics~\cite{NandyARXIV}.
Although Eq.~(\ref{Main}) holds for arbitrary $\hat{O}$, we mainly consider the special case $\hat O\!=\! J_{\alpha}$, where $J_{\alpha}\!=\!\sum_{j=1}^N \sigma_\alpha^{(j)}/2$ ($\alpha\!=\!x,y,z$) are collective spin operators of $N$ qubits, with $\sigma_\alpha^{(j)}$ being Pauli matrices.
It predicts  $\bar{F}_{\rm chaos}\!\simeq\!N$ when the chaotic dynamics extends over the full $N$-qubit space $\mathcal K\!=\!\otimes^N\mathbb C^{2}$ ($K\!=\!2^N$) or the subspace subjected to bit reversal symmetry or parity symmetry, which explains why only the linear scaling of QFI can be obtained in a chaotic Ising model.
Instead, we find $\bar{F}_{\rm chaos}\!\simeq\! N^2/3$ for the permutation-symmetric subspace $\mathcal K\!=\!{\rm Sym}^N ( \mathbb C^2)$ ($K\!=\!N\!+\!1$), as numerically confirmed by various kicked-top models. 
In this case, our results indicate that chaotic  collective spin systems~\cite{FN0} can generate chaotic permutation-symmetric $N$-qubit states which exhibit at least $\lfloor N/3\rfloor$-particle scalable entanglement. 
Furthermore, for permutation-symmetric states, we clarify that the universal scaling of QFI in Eq.~(\ref{Main}) and its independence from spin direction $\alpha$ can distinguish chaotic from integrable collective spin systems with positive Lyapunov exponents (LEs)~\cite{PappalardiPRA2018,LerosePRA2020,ChangPRA2021}, in contrast to OTOCs.

Equation~(\ref{Main}) has several remarkable consequences~\cite{note_thermalization}. 
First, it provides a unifying view of randomness and quantum chaos, as expressed by the equality
\begin{flalign}\label{Main2}
\bar{F}_{\rm chaos}[\hat O] \!\simeq\! \bar{F}_{\rm rand}[\hat O]  \equiv \int  d\mu(U) F_Q\big[U\ket{\phi_0},\hat O\big],
\end{flalign} 
up to the leading terms in $K$, where the right-hand side quantity is the QFI averaged over random  states in the space $\mathcal K$.
Reference~\cite{OszmaniecPRX2016} proved that $\bar{F}_{\rm rand}[\hat O]=4\tr_{\mathcal K}[\hat O^2]/(K+1)$ $-4(\tr_{\mathcal K}[\hat O])^2/K(K+1)$, which agrees with Eq.~(\ref{Main}) for $K\!\gg\!1$~\cite{FN-PRX}.
Notice that the unitary operators generating chaotic states are not Haar random~\cite{RobertsJHEP2017}. 
Actually, the implementation of random unitary operators in a quantum system is  exponentially hard in general~\cite{EmersonScience2003, NakataPRX2017, ChoiNATURE2023}.
Our findings demonstrate that randomness displayed by the QFI in Eq.~(\ref{Main2}) can be achieved in chaotic $N$-qubit systems involving only two-body interactions and initially polarized states, realizable in current atomic quantum simulators~\cite{ChaudhuryNATURE2009,NeillNP2016,KrithikaPRE2019,ChalopinNC2018,MourikPRE2018,BrittonNature2012,GarttnerNP2017,MarkPRL2023,BornetNATURE2023,LantanoARXIV}.

\begin{figure}[t]
\centering
\includegraphics[width=3.5in]{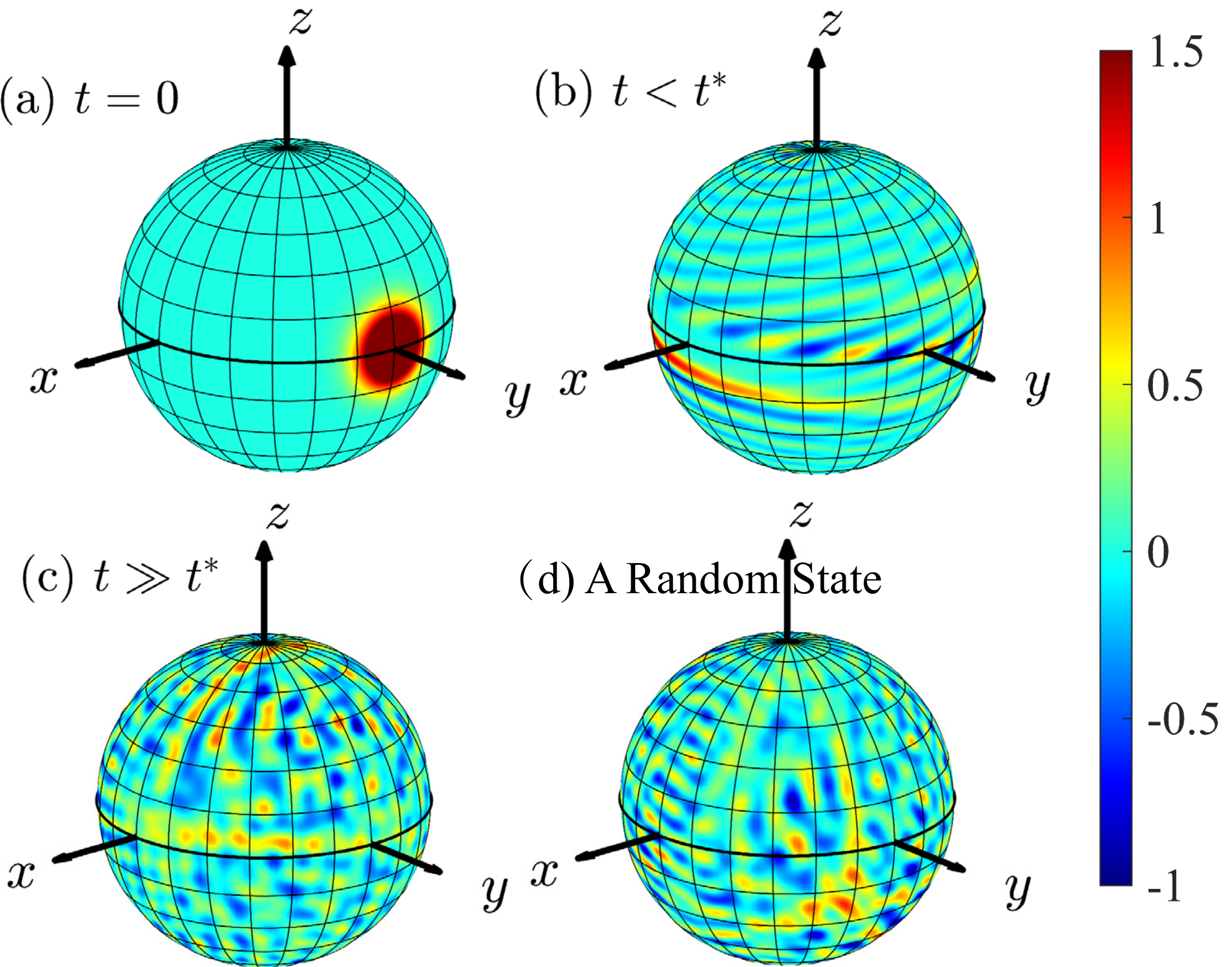}
\caption{
Panels (a)-(c) show the Wigner distributions (colormap) of the evolved state $\ket{\psi_{\rm chaos}(t)}$ for the COE chaotic model Eq.~(\ref{Floquet-COE}): (a) $t\!=\!0$, corresponding to the initial coherent spin state aligned along the $y$ axis; (b) $t\!<\!t^*$; and (c) $t\!\gg\!t^*$.
Panel (d) shows the Wigner distribution of a typical example of random state with QFI close to $\bar{F}_{\rm rand}$, Eq.~(\ref{Main2}).
}\label{figWigner}
\end{figure}

Furthermore, interesting insights can be gained by using the fundamental relation between the QFI and the fidelity OTOC, $\mathcal F_{\hat O}(\theta,t)\!\equiv\!\ex{\psi_0|W^\dag_{\hat O}(\theta,t)\rho^\dag (0)W_{\hat O}(\theta,t)\rho(0)|\psi_0}$~\cite{MacriPRA2016,GarttnerPRL2018, PappalardiPRA2018, LerosePRA2020}:
\begin{flalign}
\label{fidelity}
-2 \frac{\partial^2 \mathcal F_{\hat O}(\theta,t)}{\partial\theta^2} = F_Q[\ket{\psi_{\rm chaos}(t)},\hat O],
\end{flalign}
where $\rho(0)\!=\!\ket{\psi_0}\bra{\psi_0}$ is the initial pure state and $W_{\hat O}(\theta,t)\!=\!U_{\rm chaos}^\dag (t)\exp(i\theta \hat O)U_{\rm chaos}(t)$~\cite{FN1}.
Let us consider chaotic states in the permutation-symmetric subspace and use $\hat O\!=\! J_{\alpha}$.
In this case, 
Eqs.~(\ref{Main}) and~(\ref{fidelity}) predict that typically (namely, by replacing $F_Q[\ket{\psi_{\rm chaos}(t)},J_{\alpha}]$ for $t\!\gg\! t^*$ by its time average $\bar{F}_{\rm chaos}[J_{\alpha}]$) the fidelity decreases sharply after a collective spin rotation $\exp(i\delta\theta \hat{J}_\alpha)$ by an angle $\delta \theta\!\sim\! \sqrt{3}/N$ around any axis $\alpha$: an evident non-Gaussian feature.
Consequently, we anticipate that chaotic states exhibit uniform angular substructures in the generalized Bloch sphere, with a characteristic angular size $\delta \theta$. 
This is confirmed by plotting the Wigner distribution of chaotic states obtained from a kicked top model (as discussed below) and illustrated in Fig.~\ref{figWigner}(a)-(c).
Starting from a coherent spin state [panel (a)], the Wigner distribution of the chaotic state initially develops elongated structures for  $t\!<\!t^*$ [panel (b)].
Subsequently, for $t\!\gg\!t^*$, we observe characteristic isotropic substructures [panel (c)], as expected, which indicates a high metrological sensitivity~\cite{ZurekSCIENCE2001, ToscanoPRA2006}.
For comparison, panel (d) displays the Wigner distribution of a random state with QFI close to the average given by Eq.~(\ref{Main2}).
The similarity between the Wigner distribution of $\ket{\psi_{\rm chaos}(t\!\gg\! t^*)}$ with that of a random state visually supports the results discussed above.

In the following, we detail the derivation and consequences of Eq.~(\ref{Main}), also see Appendix, and present the results of numerical simulations.

\section{QFI for quantum chaotic dynamics}
First, let us briefly recall the fundamental relationship between QFI and entanglement~\cite{TothJPA2014, PezzeRMP2018}. 
For a generic pure $N$-qubit state $\ket{\psi} \bra{\psi}$, the QFI~\cite{BraunsteinPRL1994} is given by
\begin{flalign}
\label{QFI-1}
F_Q[\ket{\psi},\hat O]=4(\bra{\psi}\hat O^2 \ket{\psi}-\bra{\psi} \hat O \ket{\psi}^2).
\end{flalign}
Violation of the inequality
$F_Q[\ket{\psi},J_{\alpha}]\!\leq\! sk^2+r^2$, for $\hat O\!=\!J_{\alpha}$,
indicates that $\ket{\psi}$ contains at least $(k\!+\!1)$-particle entanglement among the $N$ qubits~\cite{PezzePRL2009,HyllusPRA2012,TothPRA2012}, where $s\!=\!\lfloor{N/k}\rfloor$ represents the largest integer less than or equal to $N/k$, and $r\!=\!N\!-\!sk$.
QFI can also be used to detect inseparability and entanglement depth~\cite{HongPRA2015,RenPRL2021}.
Next, we will derive the QFI for chaotic systems by computing Eq.~(\ref{QFI-1}) using RMT and the ergodicity hypothesis.

The chaotic systems under consideration exhibit energy spectrum structures that, although potentially belonging to distinct symmetry classes, are effectively modeled by Dyson's three circular ensembles of unitary random matrices~\cite{DysonMathPhys1962-1,DysonMathPhys1962-2,DysonMathPhys1962-3,MehtaBook2004,HaakeBook2010}, which validates the reliability of employing RMT.
These are the circular orthogonal ensemble (COE), the circular unitary ensemble (CUE), or the circular symplectic ensemble (CSE), each gives distinct distributions of energy level spacings
\begin{flalign}\label{LS}
    P(S)=\left\{\begin{array}{lcl}
    \frac{S\pi}{2}\exp\left(-\frac{S^2\pi}{4}\right),&&{\rm COE},\\
    \frac{S^2 32}{\pi^2}\exp\left(-\frac{S^2 4}{\pi}\right),&&{\rm CUE},\\
    \frac{S^4 2^{18}}{3^6\pi^3}\exp\left(-\frac{S^2 64}{9\pi}\right),&&{\rm CSE},
    \end{array}\right.
\end{flalign}
corresponding to linear, quadratic, and quartic level repulsion, respectively.
Here, $S_n\!=\!(E_{n+1}$ $-E_n)/\Delta E$ with the $E_n$ being the ordered eigenenergies of $H$ and $\Delta E$ the mean level spacing.

Let $\{\ket{b_m}\}_{m=1}^K$ be a basis of the Krylov space formed by eigenstates of the corresponding random matrix associated with the chaotic Hamiltonian $H$. 
We recall that the Krylov space $\mathcal K\!\equiv\! {\rm span}\{H^n\ket{\psi_0}\}_{n=0}^\infty$ is the  minimal subspace of the full Hilbert space $\mathcal H$ confining the time evolution of $\ket{\psi(t)}_{\rm chaos}\!\equiv\! U(t)\ket{\psi_0}$ where $H\!\equiv\! i\ln U(t)/t$~\cite{NandyARXIV}.
We write 
\begin{eqnarray}
\ex{\hat O^2(t)}\! = \!\sum_{m,m'=1}^K a_{m}^*(t)a_{m'}(t)\ex{b_m|\hat O^2|b_{m'}},
\end{eqnarray}
 where $a_m(t)\!=\!\ex{b_m|\psi_{\rm chaos}(t)}$, to evaluate Eq.~(\ref{QFI-1}).
According to the ergodicity hypothesis~\cite{GoldsteinEPJH2010,PolkovnikovRMP2011}, the time averaging for vector $\vect{a}(t)\!=\![a_1(t),a_2(t),\ldots, a_K(t)]$ as in 
Eq.~(\ref{Fchaos}) can be replaced by averaging over random matrices (that we indicate with an overline in the following).
We have $\overline{\ex{\hat O^2}}\!=\! \sum_{m,m'=1}^K \overline{a_{m}^*a_{m'}}\ex{b_m|\hat O^2|b_{m'}}$ and 
use the result~\cite{HaakeBook2010,UllahPR1965,DAlessioAP2016}
\begin{flalign}
\overline{a_m^* a_{m'}} = \frac{\Gamma(\beta K/2)}{\Gamma(\beta K/2+1)}\frac{\Gamma(1+\beta/2)}{\Gamma(\beta/2)}\, \delta_{m,m'},
\end{flalign}
where $\beta\!=\!1,2,4$ refers to COE, CUE, and CSE, respectively, and $\Gamma(\cdot)$ is the Gamma function.
Noticing that $\sum_m\ex{b_m|\hat O^2|b_{m}} \!=\! \tr_{\mathcal K}[\hat O^2]$, we find $\overline{\ex{\hat O^2}}\!=\!\tr_{\mathcal K}[\hat O^2]/K$,
in the limit $K\!\gg\!1$.
In the same limit,
 $\overline{\ex{O}^2} \!=\!
\left(\tr_{\mathcal K}[\hat O]\right)^2/ K^2\!+\! \mathcal O\left(\tr_{\mathcal K}[\hat O^2]/K^2\right)$, see Appendix A. 
Equation~(\ref{Main}) is recovered by substituting the above correlation functions into Eq.~(\ref{QFI-1}).
We emphasize that, in the limit $K\!\gg\! 1$, $\beta$-dependent terms do not contribute to the leading terms of QFI (see Appendix A) and Eq.~(\ref{Main}) thus universally applies to chaotic dynamics irrespective of whether they are described by COE, CUE, or CSE.

\section{Chaotic dynamics in the permutation-symmetric subspace}

We consider $N$-qubit systems with Hamiltonian involving collective spin operators, thus having   $\vect{J}^2\!=\! \sum_{\alpha=x,y,z} J_\alpha^2$ as a conserved quantity (Casimir invariant).
By taking $\ket{\psi_0}$ as a spin-polarized state, we find that the Krylov space is given by the permutation-symmetric subspace $\mathcal K\!=\!{\rm Sym}^N ( \mathbb C^2)$ of dimension $K\!=\!N\!+\!1$ with $\vect{J}^2\!=\!N(N\!+\!2)/4\mathbb I_\mathcal{K}$.
In this case, 
\begin{eqnarray}
\tr_{\mathcal K}[J_{\alpha}^2]\!=\!N(N\!+\!1)(N\!+\!2)/12, \quad \tr_{\mathcal K}[J_{\alpha}]\!=\!0,
\end{eqnarray}
 and thus Eq.~(\ref{Main}) predicts $\bar{F}_{\rm chaos}[J_{\alpha}] \simeq N^2/3$.

We validate these predictions using specific examples of chaotic kicked-top models with all-to-all spin-spin interactions and addressing the different Dyson ensembles.
Considering stroboscopic times, we write $\ket{\psi_{\rm chaos}(n)}\!=\! U^n \ket{\psi_0}$, where $U$ is a Floquet operator for the different random matrix models~\cite{HaakeBook2010,FN2}:
\begin{flalign}\label{Floquet-COE}
&U_{\rm COE}\!=\! \exp
\left(-i \frac{C}{N} J_z^2\right) \exp\left(-i A J_x\right),\\ \label{Floquet-CUE}
&U_{\rm CUE}\!=\!\exp\left(\!-\!i\frac{\lambda' J_y^2}{N}\right)
\exp\left(-i\frac{\lambda J_z^2}{N}\right)
e^{-ip\vect J_x},\\ \label{Floquet-CSE}
&U_{\rm CSE}\!=\! \exp(-iV) \exp(-iH_0),
\end{flalign}
with $H_0\!=\!2\lambda_0 J_z^2/N$ and $V\!=\!8\lambda_1 J_z^4/N^3\!+\!2\lambda_2( J_x J_z+ J_z  J_x)/N\!+\!2\lambda_3(J_x  J_y\!+\! J_y  J_x)/N$,
for COE, CUE and CSE, respectively.
%

%%%%%%%%%%%%%%%%%%%%%%%%%%%%%%%%%%%%
%%%%%%%%%%%%%%%%%%%%%%%%%%%%%%%%%%%%
%%%%%%%%%%%%%%%%%%%%%%%%%%%%%%%%%%%%

%%%%%%%%%%%%%%%%%%%%%%%%%%%%%%%%%%%%
%% FIGURE 2
%%%%%%%%%%%%%%%%%%%%%%%%%%%%%%%%%%%%
\begin{figure}[htbp]
\centering
\includegraphics[width=0.7\textwidth]{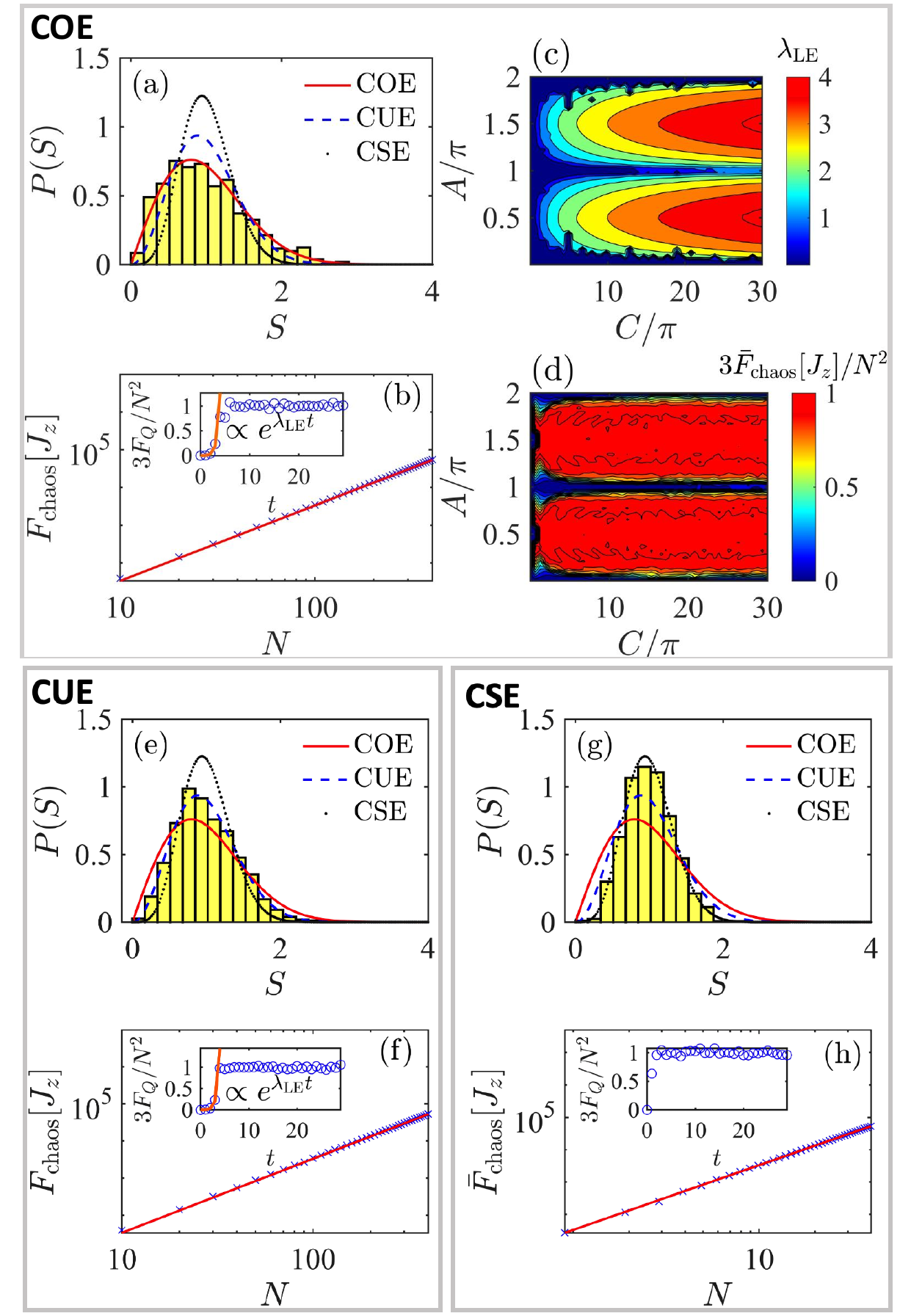}
\caption{
Level spacing distributions $P(S)$ [panels (a), (e), and (g)], LE $\lambda_{\rm LE}$ [panel (c)], and  long-time averaged QFI $\bar{F}_{\rm chaos}$ [panels (b), (d), (f) and (h)] for COE [panels (a)-(d)], CUE [panels (e)-(f)], and CSE [panels (g)-(h)] chaotic kicked top models~(\ref{Floquet-COE})-(\ref{Floquet-CSE}).
Panels (a), (e), and (g) compare numerically obtained level spacing distributions (histograms) for models Eqs.~(\ref{Floquet-COE})-(\ref{Floquet-CSE}) with analytic expressions~Eq.~(\ref{LS}) [line, dashed line, and dots, respectively], and verify their classification under RMT.
Panels (b), (f), and (h) confirm the consistency between numerically obtained $\bar{F}_{\rm chaos}$ (blue crosses) and the universal scaling $N^2/3$ of QFI from Eq.~(\ref{Main}) (red line).
Their insets depict the dynamic evolution of QFI (blue circles).
The short-time behavior is accurately described by exponential growth (red line) $\propto \exp(\lambda_{\rm LE}t)$ with $\lambda_{\rm LE}=1.874$ for $U_{\rm COE}$ and $\lambda_{\rm LE}=1.937$ for $U_{\rm CUE}$.
After the transient Ehrenfest time, the QFI exhibits barely noticeable fluctuations around the constant value given by Eq.~(\ref{Main}).
For COE model, the red region in panel (d), given by Eq.~(\ref{Main}) corresponds to the region in panel (c) with a positive LE.
In all panels, we start with a coherent spin state along the $-y$ direction and evolve it under the following chaotic conditions: $A\!=\!1.7, C\!=\!10$ for model Eq.~(\ref{Floquet-COE}), $p\!=\! 1.7,\lambda\!=\!10,\lambda'\!=\!0.5$ for model Eq.~(\ref{Floquet-CUE}), and $\lambda_0\!=\! \lambda_1\!=\!2.5,\lambda_2\!=\!5, \lambda_3\!=\! 7.5$  for model Eq.~(\ref{Floquet-CSE}).
}\label{figScaling}
\end{figure}

Figures~\ref{figScaling} provide direct numerical evidence supporting the $N^2/3$-scaling law for chaotic collective spin dynamics irrespective of the specific random matrix model.
Panels (a), (e), and (g) display numerically computed histograms of the level spacing distribution for $i\ln(U_{\rm y})$ (y=COE, CUE, CSE),  and compare them with the analytic expressions Eq.~(\ref{LS}).
These panels clearly distinguish the distinct level-spacing distributions among the three models (COE, CUE, and CSE).
The insets in panels (b), (f), and (h) plot the QFI as a function of time. 
The QFI $F_Q[\ket{\psi_{\rm chaos}(t)}, J_{z}]$ exhibits an initial exponential growth~\cite{FN3, LerosePRA2020} in a short-time regime (for $t\!\leq\! t^*$, solid red line) followed by saturation at the predicted value Eq.~(\ref{Main}), namely $\bar{F}_{\rm chaos}[J_{\alpha}]\!=\!N^2/3$. 
In panels (b), (f), and (h), we plot the QFI extracted from long-time averaging for $t\gg t^*$, as a function of the number of qubits $N$.
The QFI plotted in all figures corresponds to $J_z$. In the Appendix D, we also plot the QFI with respect to $J_x$ and $J_y$ and numerically verify that the universal scaling of QFI holds for all spin directions.
A difference between spin directions is observed only for a short time $t<t^*$, see Appendix D.
The numerical results agree very well with Eq.~(\ref{Main}) even for relatively small values of $N$.
Moreover, panel (c) shows the LE computed numerically, see Appendix B, while panel (d) plots the long-time averaged QFI, as a function of the parameters $A$ and $C$ of the model~(\ref{Floquet-COE}).
The agreement between the semi-elliptical regions with finite LEs and the region exhibiting $N^2/3$ scaling of the QFI further supports the universality of Eq.~(\ref{Main}).

%%%%%%%%%%%%%%%%%%%%%%%%%%%%%%%%%%%%
%% FIGURE 3
%%%%%%%%%%%%%%%%%%%%%%%%%%%%%%%%%%%%
\begin{figure}[t]
\centering
\includegraphics[width=0.75\columnwidth]{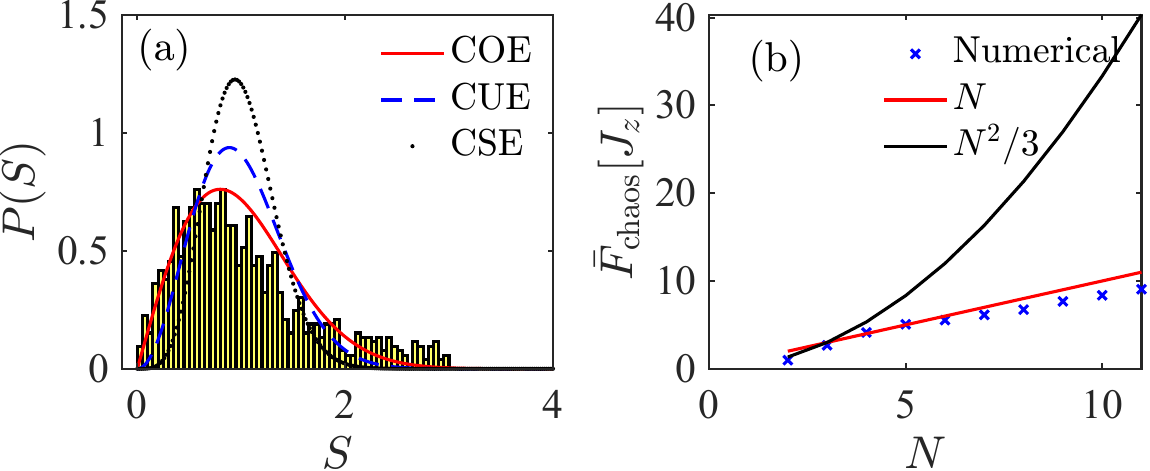}
\caption{
Level spacing distributions [panel (a)] and long-time averaged QFI [panels (b)] for the chaotic Ising model, Eq.~(\ref{cIsing}) with parameters $J\!=\!h\!=\!\lambda=1$.
}\label{figcIsing}
\end{figure}

%%%%%%%%%%%%%%%%%%%%%%%%%%%%%%%%%%%%
%%%%%%%%%%%%%%%%%%%%%%%%%%%%%%%%%%%%
%%%%%%%%%%%%%%%%%%%%%%%%%%%%%%%%%%%%

\section{Chaotic dynamics in the full Hilbert space}

Another distinct chaotic model is constructed by applying both transverse and longitudinal fields to the $N$-qubit Ising model
\begin{flalign}\label{cIsing}
H_{\rm cIsing}=\sum_{i=1}^{N}(J\sigma_{i}^{x}\sigma_{i+1}^{x}+h\sigma_{i}^{x}+\lambda\sigma_{i}^{z}),
\end{flalign}
where open boundary conditions are adopted. 
Energy-level spacing statistics, see Fig.~\ref{figcIsing}(a),
indicate that this model is chaotic for $J\!=\!h\!=\!\lambda=1$~\cite{NohPRE2021,KarthikPRA2007}.
We argue that the dynamics extend over the Krylov space $\mathcal K\!=\!\otimes^N\mathbb C^2$, which coincides with the full $N$-qubit Hilbert space of dimension $K=2^N$.
In this case, $\tr_{\mathcal K}[J_{\alpha}^2]\!=\!N2^{N-2}$, $\tr_{\mathcal K}[J_{\alpha}]\!=\!0$,  and thus Eq.~(\ref{Main}) implies $\bar{F}_{\rm chaos}[J_{\alpha}]\!\simeq\!N$. 
This result holds even in the presence of  bit reversal symmetry or parity symmetry, which will further reduce the full $N$-qubit Hilbert space, see Appendix A1.
Our prediction is confirmed by numerical analysis in Fig.~\ref{figcIsing}(b), covering system sizes from $N=2,3,\ldots,11$.
Even for moderately small values of $K$, our theoretical result remains valid.
More generally, Eq.~(\ref{Main}) suggests an operational approach to identify the Krylov space dimension~\cite{XiaARXIV2024} for chaotic dynamics with variable-range interaction, to be explored in future research.

\section{Comparison with quantum systems with semiclassical correspondence}
The QFI of quantum systems with semiclassical correspondence grows exponentially at a rate given by the largest LE, including both chaotic and integrable models~\cite{LerosePRA2020}.
Some integrable models~\cite{CameoPRE2020, HummelPRL2019} exhibit positive LEs at unstable points, leading to false detection of quantum chaos using OTOCs. 
However, these models do not conform to the RMT description, indicating they are not truly chaotic and thus invalidating Eq~(\ref{Main}). 
Therefore, compared with OTOCs, our universal scaling Eq.~(\ref{Main}), better excludes these atypical integrable models with positive LEs from chaotic models, providing a clearer indicator of quantum chaos.

As an illustration, we examine the Lipkin-Meshkov-Glick (LMG) model \cite{LipkinNuclearP1965,MeshkovNuclearP1965,GlickNuclearP1965}, 
\begin{eqnarray}
H_{\rm LMG} = \Omega  J_z -\frac{2\xi}{N} J_x^2,
\end{eqnarray}
which is characterized by the generalized Gaudin Lie algebra, enabling exact solvability through the Bethe ansatz \cite{OrtizNPB2005}.
It finds experimental realizations in various atomic ensebles~\cite{PezzeRMP2018,StrobelSCIENCE2014, MuesselPRA2015,BohnetSCIENCE2016} and nuclear magnetic resonance platforms~\cite{FerreiraPRA2013}.
Due to the existence of an unstable point, this integrable model has a positive LE $\lambda_{\rm LE} \!=\! \sqrt{\Omega(2\xi-\Omega)}$ in the phase of $\Omega(\Omega\!-\!2\xi)\!<\!0$~\cite{CameoPRE2020}, also see Appendix C.
Figure~\ref{figLMG}(a) plots the evolution of the QFI $F_Q[\ket{\psi(t)}, J_{\alpha}]$ of an initial coherent spin state prepared at the semiclassical unstable fixed point, and for the given $\alpha\!=\!z$.
Initially, the QFI exhibits exponential growth $\!\sim\! e^{4 \lambda_{\rm LE} t}$, resembling chaotic dynamics~\cite{LerosePRA2020, ChangPRA2021}.
However, the saturation value reached for time $t\gg t^*$ does not agree with the value $N^2/3$, Eq.(\ref{Main}), of truly quantum chaotic systems, despite the positive LE. 
Furthermore, in contrast to Eq.~(\ref{Main}) the QFI of the unstable integrable model shows a strong dependence in $\alpha$, see Appendix D.

%

%%%%%%%%%%%%%%%%%%%%%%%%%%%%%%%%%%%%
%% FIGURE 4
%%%%%%%%%%%%%%%%%%%%%%%%%%%%%%%%%%%%
\begin{figure}[t]
\centering
\includegraphics[width=0.75\columnwidth]{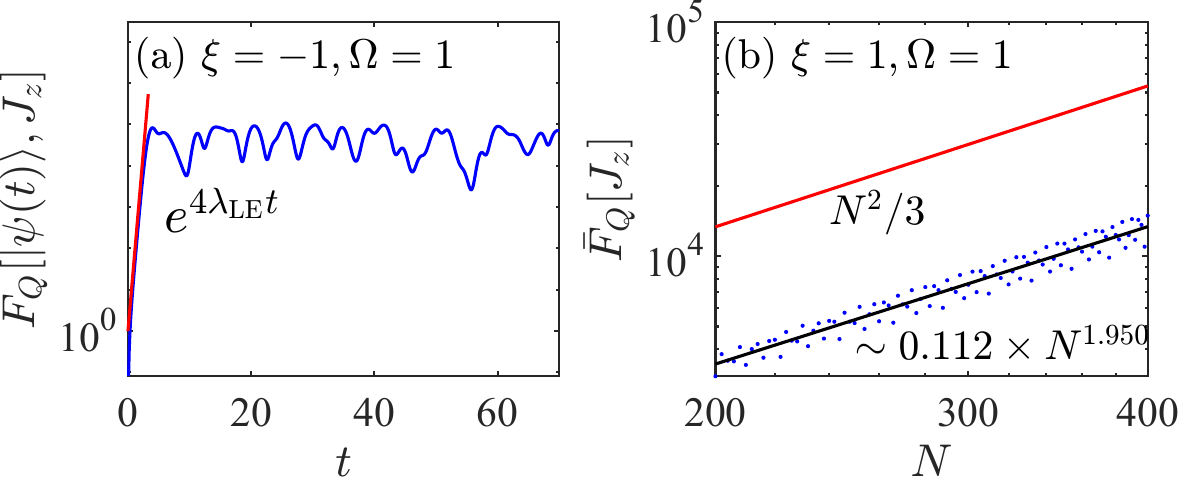}
\caption{
Time-evolution of QFI (a) and long-time averaged QFI (b) in the integrable LMG model for the initial unstable coherent spin state along the $-z$ direction.
The red straight line in (a) (semi-logarithmic coordinate) corresponds to the exponential curve with the rate given by four times LE, $\exp(4\lambda_{\rm LE}t)$.
Blue dots in (b) are numerical results, which are fitted by a black solid line with scaling of $F_Q\simeq0.112N^{1.950}$. This behavior deviates from the universal scaling for chaotic collective spin models $N^2/3$, as marked by the solid red line.
}\label{figLMG}
\end{figure}
%%%%%%%%%%%%%%%%%%%%%%%%%%%%%%%%%%%%
%%%%%%%%%%%%%%%%%%%%%%%%%%%%%%%%%%%%
%%%%%%%%%%%%%%%%%%%%%%%%%%%%%%%%%%%%

\section{Conclusion and discussion}
Through QFI, we have provided a unifying approach to randomness, multipartite entanglement, and quantum chaos, with a fundamental dependence on the dimension of the Krylov space or symmetry.
Our analysis has been restricted to the broad class of chaotic systems governed by Dyson's three random matrix ensembles,
with a focus on the long-time dynamical behavior of the QFI.
Nevertheless, we expect possible generalizations beyond the threefold  class~\cite{ZirnbauerJMP1996,AltlandPRB1997,AltlandPReport2002,ZirnbauerArXiv2004,HeinznerCMP2005} and open chaotic systems~\cite{KawabataPRXQ2023}, as well as a dependence on possible deviations from RMT statistics.
It is important to emphasize that, while fast scrambling is often viewed as a necessary condition for the rapid buildup of entanglement, it is not sufficient to achieve scalable entanglement from separable initial states.
For instance, despite exhibiting fast scrambling, the chaotic Ising model fails to dynamically generate Heisenberg-scaling (HS) entangled states due to the absence of an appropriate symmetry structure.
From the perspective of quantum metrology, our results demonstrate the remarkable capability of chaotic collective spin systems with SU(2) symmetry to generate scalable multipartite entanglement at an exponentially fast rate.
In particular, the QFI exhibits a universal scaling
$\bar{F}_{\rm chaos}[J_\alpha]=N^2/3$, indicating the emergence of scalable  multipartite entanglement~\cite{AugusiakPRA2012}.
Unlike GHZ or squeezed states, which achieve large QFI only for specific axes~\cite{PezzeRMP2018}, the entanglement generated by chaotic collective spin dynamics is axis-independent and accessible from any symmetric separable initial state, in contrast also to entanglement from integrable dynamics such as one-axis twisting~\cite{KitagawaPRA1993}.

Our findings offer a complementary perspective to previous work.
Ref.\cite{ChuPRL2023} established lower bounds on the time required to prepare HS-entangled states based on Lieb-Robinson constraints, assuming the existence of such target states.
Ref.\cite{ShiPRL2024} showed that under short-range local Hamiltonians, the QFI from separable initial states scales only linearly with system size, ruling out generic HS generation.
Ref.~\cite{ChuArXiv2024} highlighted the role of correlation spreading in fast entanglement growth, but without addressing symmetry constraints.
In contrast, our results demonstrate that symmetry is a crucial enabling factor for HS scaling in chaotic dynamics, going beyond fast scrambling alone.
In addition, while quantities such as OTOCs primarily capture short-time behavior, the QFI offers a complementary probe of long-time quantum dynamics.
This enables a clear distinction between fast-scrambling integrable systems and genuinely chaotic ones—a distinction that remains elusive when using short-time diagnostics alone.
Altogether, our work deepens the understanding of dynamical entanglement generation and its fundamental limits, with significant implications for quantum metrology and quantum information processing in quantum many-body systems.

\section*{Acknowledgements}
We would like to thank Satoya Imai, Li Gan, Silvia Pappalardi, Jing Yang, Sandro Wimberger and Wei Xia for discussions.

% TODO: include funding information
\paragraph{Funding information}
This work was supported by the European Commission through the H2020 QuantERA ERA-NET Cofund in Quantum Technologies project ``MENTA'' and received funding under Horizon Europe programme HORIZONCL4-2022-QUANTUM-02-SGA via the project 101113690
(PASQuanS2.1)

\begin{appendix}\label{appendix A}
\numberwithin{equation}{section}

\section{Derivation of long-time averaged QFI for chaotic systems, Eq.~(2)}

Let $\{\ket{b_m}\}_{m=1}^K$ be a basis of the Krylov space $\mathcal K\!\equiv\! {\rm span}\{H^n\ket{\psi_0}\}_{n=0}^\infty$formed by eigenstates of the corresponding random matrix associated with the chaotic Hamiltonian $H$.
The evolved chaotic state $\ket{\psi_{\rm chaos}(t)}=e^{-i Ht}\ket{\psi_0}$  can be expanded in the Krylov space as 
\begin{eqnarray}
    \ket{\psi_{\rm chaos}(t)}=\sum_{m=1}^K a_m(t)\ket{b_m},
\end{eqnarray}
where $a_m(t)=\ex{b_m|\psi_{\rm chaos}(t)}$.
We want to compute 
\begin{flalign} %\label{Fchaos}
\bar{F}_{\rm chaos}[\hat O]\equiv \lim_{T\to\infty} \frac{1}{T-t^*}\int_{t^*}^{T} dt F_Q[\ket{\psi_{\rm chaos}(t)}, \hat O],
\end{flalign}
where $\hat O$ is a Hermitian operator and
\begin{flalign}
F_Q[\ket{\psi_{\rm chaos}(t)}, \hat O]=4 \big( \ex{ \hat O^2(t)} - \ex{\hat O(t)}^2 \big)
\end{flalign}
is the QFI of the evolved chaotic state $\ket{\psi_{\rm chaos}(t)}$, with
\begin{eqnarray}
&&\ex{\hat O^\kappa(t)}=\sum_{m,m'=1}^K a_{m}^*(t)a_{m'}(t)\ex{b_m|\hat O^\kappa|b_{m'}}.  
\end{eqnarray}
Combining the above equations we have
\begin{eqnarray} \label{Fishertimeaverage}
&&\bar{F}_{\rm chaos}[\hat O]\n 
& &=  
4 \sum_{m,m'=1}^K \ex{b_m|\hat O^2|b_{m'}} 
\bigg( \lim_{T\to\infty} \frac{1}{T-t^*}\int_{t^*}^{T} dt \, a_{m}^*(t)a_{m'}(t) \bigg) \\
& & - 4 \sum_{m_1,m_1',m_2,m_2'=1}^K \ex{b_{m_1}|\hat O|b_{m_1'}}\ex{b_{m_2}|\hat O|b_{m_2'}} \bigg( \lim_{T\to\infty} \frac{1}{T-t^*}\int_{t^*}^{T} dt \, a_{m_1}^*(t)a_{m_1'}(t)
a_{m_2}^*(t)a_{m_2'}(t) \bigg).\nonumber
\end{eqnarray}
According to the ergodicity hypothesis~\cite{GoldsteinEPJH2010,PolkovnikovRMP2011}, the time averaging for vector $\vect{a}(t)\!=\![a_1(t),a_2(t),$ $\ldots, a_K(t)]$ can be replaced by averaging over random matrices (that we indicate with an overline in the following), namely
\begin{eqnarray}
&&\lim_{T\to\infty} \frac{1}{T-t^*}\int_{t^*}^{T} dt \, a_{m}^*(t)a_{m'}(t) = \overline{a_{m}^*(0)a_{m'}(0)}, \n
&&\lim_{T\to\infty} \frac{1}{T-t^*}\int_{t^*}^{T} dt \, a_{m_1}^*(t)a_{m_1'}(t)
a_{m_2}^*(t)a_{m_2'}(t) = \overline{a_{m_1}^*(0)a_{m_1'}(0)a_{m_2}^*(0)a_{m_2'}(0)},
\end{eqnarray}
where $a_m(0)=\ex{b_m|\psi_0}$ is the projection of the initial state $\ket{\psi_0}$ on the eigenvector $\ket{b_m}$, with  $\ket{b_m}$ being the $m$-th eigenvector of one of the random matrices in the random matrix ensemble.
We can define another basis formed by $\{\ket{\psi_0}, \ket{\psi_1},\ldots,\ket{\psi_{K-1}}\}$ and define $a_m\equiv \ex{b_1|\psi_{m-1}}$ as the $m$-th element of the eigenvector $\ket{b_1}$.
Random matrix theory (RMT) provides~\cite{HaakeBook2010,UllahPR1965,DAlessioAP2016}
\begin{equation}
 \overline{a_{m}(0)^*a_{m'}(0)} =
  \overline{a_{m}^*a_{m'}}
  =\overline{|a_{m}|^2}\, \delta_{m,m'}   
\end{equation}
and 
\begin{eqnarray} &&\overline{a_{m_1}^*(0)a_{m_1'}(0)a_{m_2}^*(0)a_{m_2'}(0)} 
= 
\overline{a_{m_1}^*a_{m_1'}a_{m_2}^*a_{m_2'}} \n
& &=\overline{\vert a_{m} \vert^4}\, \delta_{m_1,m}\delta_{m_1',m}\delta_{m_2,m}\delta_{m_2',m} +
 \overline{\vert a_{m_1} \vert^2 \vert a_{m_2} \vert^2}\,\big( \delta_{m_1,m'_1}\delta_{m_2,m'_2} + \delta_{m_1,m'_2}\delta_{m_2,m'_1} \big).
\end{eqnarray}
where
\begin{eqnarray}\label{RMT-average}
\overbar{|a_1|^{2m_1}|a_2|^{2m_2}\cdots |a_K|^{2m_K}}
&=&\frac{\int \left(\prod_{i=1}^K|a_i|^{2m_i}\right)\delta\left(\sum_{i=1}^K |a_i|^2-1\right)\left(\prod_{i=1}^K|a_i|^{\beta-1} {\rm d}|a_i|\right)}{\int\delta\left(\sum_{i=1}^K |a_i|^2-1\right)\left(\prod_{i=1}^K {\rm d}|a_i|\right)}\n
&=&\frac{\Gamma(\beta K/2)}{\Gamma(\beta K/2+\sum_{i=1}^K m_i)}\frac{\prod_{i=1}^K\Gamma(m_i+\beta/2)}{\prod_{i=1}^K\Gamma(\beta/2)}.
\end{eqnarray} 
Here,  $m_i$ are non-negative integers, $\Gamma(\cdot)$ is the Gamma function, and $\beta\!=\!1,2,4$ refers to COE, CUE and CSE, respectively.  
Equation~\ref{Fishertimeaverage} rewrites as 
\begin{equation} \label{Fchaos_supp}
\bar{F}_{\rm chaos}[\hat O] = 4(\overline{\ex{\hat O^2}}- \overline{\ex{\hat O}^2}),    
\end{equation}
with 
\begin{subequations}\label{CorFun}
\begin{eqnarray}
\overline{\ex{\hat O^2_\alpha}} &=& \sum_{m=1}^K \overline{|a_{m}|^2}\ex{b_m|\hat O^2|b_{m}},\\
\overline{\ex{\hat O}^2} &=& \sum_{m=1}^K \overline{|a_{m}|^4}\ex{b_{m}|\hat O|b_{m}}^2\\ 
& &\!+\!
\sum_{m_1\neq m_2}^K
\overline{|a_{m_1}|^2|a_{m_2}|^2}
[\ex{b_{m_1}|\hat O|b_{m_1}}\ex{b_{m_2}|\hat O|b_{m_2}}
\!+\!\ex{b_{m_1}|\hat O|b_{m_2}}\ex{b_{m_2}|\hat O|b_{m_1}}], \nonumber
\end{eqnarray}    
\end{subequations}
where the second term is obtained by considering three kinds of nonzeros contributions in Eq.~(\ref{Fishertimeaverage}): $m_1\!=\!m_1'\!=\!m_2\!=\!m_2'$, $m_1\!=\!m_1'\neq m_2\!=\!m_2'$, and $m_1\!=\!m_2'\neq m_2\!=\!m_1'$.
From Eq.~(\ref{RMT-average}), we have
\begin{subequations}\label{RMT-average-2}
\begin{eqnarray}
&&\overbar{|a_m|^2}=\frac{\Gamma(\beta K/2)\Gamma(1+\beta/2)}{\Gamma(\beta K/2+1)\Gamma(\beta/2)}=\frac{1}{K}, \quad  \forall m \\ 
&&\overbar{|a_m|^4}=\frac{\Gamma(\beta K/2)\Gamma(2+\beta/2)}{\Gamma(\beta K/2+2)\Gamma(\beta/2)}
=\frac{\beta+2}{(\beta K+2) K}, \quad \forall m \\
&&\overbar{|a_{m_1}|^2|a_{m_2}|^2}=\frac{\Gamma(\beta K/2)\Gamma(1+\beta/2)\Gamma(1+\beta/2)}{\Gamma(\beta K/2+2)\Gamma(\beta/2)\Gamma(\beta/2)}
=\frac{\beta}{(\beta K+2)K}, \quad \forall m_1 \neq m_2.
\end{eqnarray}
\end{subequations}
Substituting Eq.~(\ref{RMT-average-2}) into Eq.~(\ref{CorFun}),
 we obtain 
\begin{eqnarray}\label{CorFun-2}
\overline{\ex{\hat O^2}} &=& \frac{1}{K}\sum_{m=1}^K \ex{b_m|\hat O^2|b_{m}},\\
\overline{\ex{\hat O}^2} &=& \frac{\beta+2}{(\beta K+2) K}\sum_{m=1}^K \ex{b_{m}|\hat O|b_{m}}^2 %\n 
+ %& &+
\frac{\beta}{(\beta K+2)K}
\sum_{m_1\neq m_2}^K
[\ex{b_{m_1}|\hat O|b_{m_1}}\ex{b_{m_2}|\hat O|b_{m_2}}\n 
& &\!+\!\ex{b_{m_1}|\hat O|b_{m_2}}\ex{b_{m_2}|\hat O|b_{m_1}}],
\end{eqnarray}
Due to $\sum_{m=1}^K \ex{b_m|\hat O^2|b_{m}}=\tr_{\mathcal K}[\hat O^2]$, the first correlation function in Eq.~(\ref{CorFun-2}) becomes 
\begin{eqnarray}\label{Main-Cor-1}
\overline{\ex{\hat O^2}}&=& \frac{\tr_{\mathcal K}[\hat O^2]}{K},
\end{eqnarray}
which is the Eq.~(8) in the Main text.
The second correlation function in Eq.~(\ref{CorFun-2}) can be evaluated as follows
\begin{eqnarray}\label{COrFUn3} 
\overline{\ex{\hat O}^2}&=& \frac{\beta}{(\beta K+2) K}\sum_{m=1}^K \ex{b_{m}|\hat O|b_{m}}^2 %\n
+ %& &+
\frac{\beta}{(\beta K+2)K}
\sum_{m_1, m_2=1}^K
[\ex{b_{m_1}|\hat O|b_{m_1}}\ex{b_{m_2}|\hat O|b_{m_2}}\n 
& &\!+\!\ex{b_{m_1}|\hat O|b_{m_2}}\ex{b_{m_2}|\hat O|b_{m_1}}] \n 
&=& \frac{\beta}{(\beta K+2) K}\sum_{m=1}^K \ex{b_{m}|\hat O|b_{m}}^2
+
\frac{\beta}{(\beta K+2)K}
\left(\sum_{m=1}^K
[\ex{b_{m}|\hat O|b_{m}}\right)^2 \n 
& &+
\frac{\beta}{(\beta K+2)K}
\sum_{m_1, m_2=1}^K
\ex{b_{m_1}|\hat O|b_{m_2}}\ex{b_{m_2}|\hat O|b_{m_1}}\n 
&=& \frac{\beta}{(\beta K+2) K}\sum_{m=1}^K \ex{b_{m}|\hat O|b_{m}}^2
+
\frac{\beta\left(\tr_{\mathcal K}[\hat O]\right)^2}{(\beta K+2)K}
+
\frac{\beta\tr_{\mathcal K}[\hat O^2]}{(\beta K+2)K},
\end{eqnarray}
where in deriving the last equation we have used the identity relation $\sum_{m=1}^K\ket{b_m}\bra{b_m}={\rm Id_{\mathcal K}}$.
Let us analyze the three terms in Eq.~(\ref{COrFUn3}) and compare them to Eq.~(\ref{Main-Cor-1}).
Let's start with the first term in Eq.~(\ref{COrFUn3}).
We use $\ex{b_{m}|\hat O|b_{m}}^2 \leq \ex{b_{m}|\hat O ^\dag\hat O |b_{m}}=\ex{b_{m}|\hat O^2|b_{m}}$, for all $\ket{b_m}$, which follows from the Hermiticity of operator $\hat O$.
This gives
\begin{equation}
\frac{\beta}{(\beta K+2) K}\sum_{m=1}^K \ex{b_{m}|\hat O|b_{m}}^2 \leq  
\frac{\beta}{(\beta K+2) K}\sum_{m=1}^K \ex{b_{m}|\hat O^2|b_{m}} = 
\frac{\beta K}{(\beta K+2)} \frac{\tr_{\mathcal K}[\hat O^2]}{K^2} \leq \frac{\tr_{\mathcal K}[\hat O^2]}{K^2}.
\end{equation}
Similarly, the last term in Eq.~(\ref{COrFUn3}) is 
\begin{equation}
\frac{\beta \tr_{\mathcal K}[\hat O^2]}{(\beta K+2)K} = \frac{\beta K}{\beta K +2} \frac{\tr_{\mathcal K}[\hat O^2]}{K^2} \leq \frac{\tr_{\mathcal K}[\hat O^2]}{K^2}.  
\end{equation}
Finally, we will show that  the second term in Eq.~(\ref{COrFUn3}) is the same or lesser order term than $\tr_{\mathcal K}[\hat O^2
]/K$, and thus it should be kept.
By writing the operator $\hat O$ in the spectral decomposition form within the Krylov space:  $\hat O=\sum_{m=1}^K\lambda_m\ket{d_m}\bra{d_m}$.
It follows that 
\begin{eqnarray}\label{Spec-decom}
    \tr_{\mathcal K}[\hat O^2]=\sum_{m=1}^K \lambda_m^2, \quad 
    (\tr_{\mathcal K}[\hat O])^2=\left(\sum_{m=1}^K\lambda_m\right)^2.
\end{eqnarray}
Applying Jensen's inequality for the convex function $f(x)=x^2$, we obtain 
\begin{eqnarray}\label{Jensen inq}
\frac{1}{K}\sum_{m=1}^K\lambda_m^2\geq \left(\frac{1}{K}\sum_{m=1}^K\lambda_m\right)^2
\end{eqnarray}
Combining Eqs.~(\ref{Spec-decom}) and (\ref{Jensen inq}), we deduce that 
\begin{eqnarray}\label{Aux-2}
    K\tr_{\mathcal K}[\hat O^2]\geq (\tr_{\mathcal K}[\hat O])^2. 
\end{eqnarray}
Thus, the second term in Eq.~(\ref{COrFUn3}) is 
\begin{equation}
\frac{\beta \left(\tr_{\mathcal K}[\hat O]\right)^2}{(\beta K+2)K} \leq  \frac{\beta }{\beta K +2} \tr_{\mathcal K}[\hat O^2] \leq \frac{\tr_{\mathcal K}[\hat O^2]}{K},
\end{equation}
which implies that this term $\beta \left(\tr_{\mathcal K}[\hat O]\right)^2/(\beta K+2)K$ is the same or lesser order term than $\tr_{\mathcal K}[\hat O^2
]/K$.
The following examples will show that $\beta \left(\tr_{\mathcal K}[\hat O]\right)^2/(\beta K+2)K$ can be on the same order of $\tr_{\mathcal K}[\hat O^2
]/K$ and thus it will contribute to the leading terms of QFI for some special situations.
Based on the above analysis,  Eq.~(\ref{COrFUn3}) can be written as
\begin{equation} \label{Main-Cor-22}
\overline{\ex{\hat O}^2} =\frac{\beta\left(\tr_{\mathcal K}[\hat O]\right)^2}{(\beta K+2)K}+ \mathcal O\left(\frac{\tr_{\mathcal K}[\hat O^2]}{K^2}\right).
\end{equation}
To analyze the influence of $\beta$, we can expand $\beta/(\beta K+2)K$ by using Taylor expansion up to $\beta$, i.e., 
\begin{eqnarray}
    \frac{\beta}{K(\beta K+2)}&=&\frac{1}{K^2}\left(1+\frac{2}{\beta K}\right)^{-1}=\frac{1}{K^2}-\frac{2}{\beta K^3}+\mathcal O (\frac{1}{K^4}).
\end{eqnarray}
Therefore, Eq.~(\ref{Main-Cor-22}) becomes 
\begin{eqnarray}\label{Main-Cor-2}
    \overline{\ex{\hat O}^2} &=&\frac{\left(\tr_{\mathcal K}[\hat O]\right)^2}{K^2}-\frac{2\left(\tr_{\mathcal K}[\hat O]\right)^2}{\beta K^3}+ \mathcal O\left(\frac{\tr_{\mathcal K}[\hat O^2]}{K^2}\right)\n    &=&\frac{\left(\tr_{\mathcal K}[\hat O]\right)^2}{K^2}+ \mathcal O\left(\frac{\tr_{\mathcal K}[\hat O^2]}{K^2}\right),
\end{eqnarray}
where the $\beta$-dependent term $\left(\tr_{\mathcal K}[\hat O]\right)^2/\beta K^3\leq \tr_{\mathcal K}[\hat O^2]/\beta K^2$ cannot contribute to the leading terms.

Combining Eqs.~(\ref{Fchaos_supp}), (\ref{Main-Cor-1}) and (\ref{Main-Cor-2}), we find 
\begin{eqnarray}\label{SM-Main}
\bar{F}_{\rm chaos}[\hat O]\!=\! 
\frac{4\tr_{\mathcal K}[\hat O^2]}{K}
\!-\!
\frac{4\left(\tr_{\mathcal K}[\hat O]\right)^2}{K^2}
\!+\!
\mathcal O\left(\frac{\tr_{\mathcal K}[\hat O^2]}{K^2}\right),
\end{eqnarray}
which is one of the main results of our manuscript, Eq.~(2) in the main text. 
Next, we will discuss different behaviors of QFI by considering two kinds of Krylov spaces.

%%%%%%%%%%%%%%%%%%%%%%%%%%%%%
\subsection{Full Hilbert space with or without bit reversal symmetry or parity symmetry }
For a fix collective spin operator $\vect J_\alpha$, we can define a basis $\{\ket{j_1,j_2,\cdots,j_N}\}$, $(j_\ell\!\in\!\{0,1\},\ell\!=\!1,2,$ $\ldots,N$) when considering the full Hilbert space $\otimes^N\mathbb C^2 $ as the Krylov space, denoted as $\mathcal K_f$.
Clearly, the dimension of $\mathcal K_f$ is $K=2^N$.
By noticing that $\ket{j_1,j_2,\cdots,j_N}$ is an eigenvector of $\vect J_\alpha$ with an eigenvalue $\sum_{\ell=1}^N(2j_\ell-1)/2$, then we have
\begin{eqnarray}\label{au4}
&&\tr_{\mathcal K_f}[\vect J_\alpha]=\sum_{m=0}^N \left(m-\frac{N}{2}\right){N\choose m}=0,\, \tr_{\mathcal K_f}[\vect J_\alpha^2]=\sum_{m=0}^N \left(m-\frac{N}{2}\right)^2{N\choose m}=2^N\frac{N}{4},
\end{eqnarray}
where we have used the results
\begin{eqnarray}
    \sum_{m=0}^N {N\choose m}=2^N, \quad  \sum_{m=0}^N m{N\choose m} =2^N\frac{N}{2}, \quad \sum_{m=0}^N m^2{N\choose m} =2^N\frac{N(N+1)}{4}.
\end{eqnarray}
Therefore, by Eq.~(\ref{SM-Main}), we find $\bar{F}_{\rm chaos}[\vect J_\alpha]=N+\mathcal O(N/2^N)$ for the full Hilbert space.

In the following we want to show that $\bar{F}_{\rm chaos}[\vect J_\alpha]\simeq N$ for the full Hilbert space with or without bit reversal symmetry or parity symmetry.
Let us first include parity symmetry.
This separates the full Hilbert space into two subspaces constructed by states with an odd number of up spins or an even number of up spins.
For the parity-odd (parity-even) subspace as the Krylov space, denoted as $\mathcal K_{+}$ ($\mathcal K_{-}$), the basis will be $\{\ket{j_1,j_2,\cdots,j_K}\}$ with $\sum_{\ell=1}^Nj_\ell$ being odd (even).
Therefore, we have 
\begin{eqnarray}\label{Cor-parity}
&&\tr_{\mathcal K_\pm}[\vect J_\alpha]=\sum_{m=0}^{\lfloor N/2\rfloor} \left(2m+\frac{1\pm1}{2}-\frac{N}{2}\right){N\choose 2m}
=2^N \frac{1\pm1}{4},\n 
&&\tr_{\mathcal K_\pm}[\vect J_\alpha^2]=\sum_{m=0}^{\lfloor N/2\rfloor}  \left(2m+\frac{1\pm1}{2}-\frac{N}{2}\right)^2{N\choose 2m}=2^N\frac{N+(1\pm1)^2}{8},
\end{eqnarray}
where we have used the results
\begin{eqnarray}
    \sum_{m=0}^{\lfloor N/2\rfloor} {N\choose 2m}=2^N\frac{1}{2}, 
    \quad
    \sum_{m=0}^{\lfloor N/2\rfloor} m{N\choose 2m}=2^{N}\frac{N}{8},
    \,
     \sum_{m=0}^{\lfloor N/2\rfloor} m^2{N\choose 2m}=2^{N}\frac{N(N+1)}{32},\, \forall\ N\geq3.
\end{eqnarray}
The dimensions of $\mathcal K_\pm$ are both $2^{N-1}$.
Therefore, substituting Eq.~(\ref{Cor-parity}) into Eq.~(\ref{SM-Main}), we conclude that $\bar{F}_{\rm chaos}[\vect J_\alpha]=N+\mathcal O(N/2^N)$ for both spaces $\mathcal K_\pm$.

Finally, we consider a full Hilbert space subjected to the bit reversal symmetry $B$.
The bit reversal operator $B$ is defined by 
\begin{eqnarray}
B\ket{j_1,j_2,\ldots,j_N}=\ket{j_N,j_{N-1},\ldots,j_1}.
\end{eqnarray}
According to the bit reversal symmetry, the separated subspaces are denoted by $\mathcal K_{b\pm}$ corresponding to $B=\pm 1$.
For simplification, we only consider the case of $  N$ being even.
The dimension of $\mathcal K_{b-}$ is half the number of
nonpalindromic binary words of length $N$, i.e., $2^{N-1}-2^{N/2-1}$, and the dimension of $\mathcal K_{b+}$ is thus $2^{N-1}+2^{N/2-1}$.
Introducing the space $\mathcal P\equiv{\rm span}\{\ket{j_1,j_2,\cdots,j_N}:$ $ B\ket{j_1,j_2,\cdots,j_N}=\ket{j_1,j_2,\cdots,j_N}\}$. 
It is straightforward to prove that 
\begin{eqnarray}\label{au5}
    \tr_{\mathcal K_{b+}}[\vect J_\alpha^\eta]=\tr_{\mathcal P}[\vect J_\alpha^\eta]+\frac{\tr_{\mathcal K_f}[\vect J_\alpha^\eta]-\tr_{\mathcal P}[\vect J_\alpha^\eta]}{2},
    \quad
      \tr_{\mathcal K_{b-}}[\vect J_\alpha^\eta]=\frac{\tr_{\mathcal K_f}[\vect J_\alpha^\eta]-\tr_{\mathcal P}[\vect J_\alpha^\eta]}{2},
\end{eqnarray}
where $\eta=1,2$.
An direction calculation gives 
\begin{eqnarray}\label{Au3}
    \tr_{\mathcal P}[\vect J_\alpha]=\sum_{m=0}^{N/2}\left(2m-\frac{N}{2}\right){N/2\choose m}=0,
    \,
      \tr_{\mathcal P}[\vect J_\alpha^2]=\sum_{m=0}^{N/2}\left(2m-\frac{N}{2}\right)^2{N/2\choose m}=2^{N/2}\frac{N}{2}.
\end{eqnarray}
Substituting Eqs.~(\ref{Au3}) and (\ref{au4}) into Eq.~(\ref{au5}) we have
\begin{eqnarray}
    \tr_{\mathcal K_{b\pm}}[\vect J_\alpha]=0,
    \quad 
      \tr_{\mathcal K_{b\pm}}[\vect J_\alpha^2]=2^N\frac{N}{8}\pm 2^{N/2} \frac{N}{4}.
\end{eqnarray}
Therefore, further by Eq.~(\ref{SM-Main}) we conclude that $\bar{F}_{\rm chaos}[\vect J_\alpha]=N+\mathcal O(N/2^{N})$ for both spaces $\mathcal K_{b\pm}$.

\subsection{Permutation-symmetric subspace}
Now we consider the permutation-symmetric subspace as the Krylov space, denoted by $\mathcal K_s$.
For a fixed spin operator $\vect J_\alpha$, we can construct a basis for $\mathcal K_s$ by using the Dickes state $\{\ket{m}\}_{m=-N/2}^{N/2}$ and meanwhile ensure $\vect J_\alpha\ket{m}=m\ket{m}$.
It finds that the dimension of $\mathcal K_s$ is $N+1$ and 
\begin{eqnarray}
&&\tr_{\mathcal K_s}[\vect J_\alpha]=\sum_{m=-N/2}^{N/2} m=0,
\quad 
\tr_{\mathcal K_s}[\vect J_\alpha^2]=\sum_{m=-N/2}^{N/2} m^2=\frac{N(N+1)(N+2)}{12}.
\end{eqnarray}
Then by using Eq.(\ref{SM-Main}), we find that  $\bar{F}_{\rm chaos}[\vect J_\alpha]=N^2/3+\mathcal O(N)$. %

\begin{figure}[t]
\centering
\includegraphics[width=0.4\columnwidth]{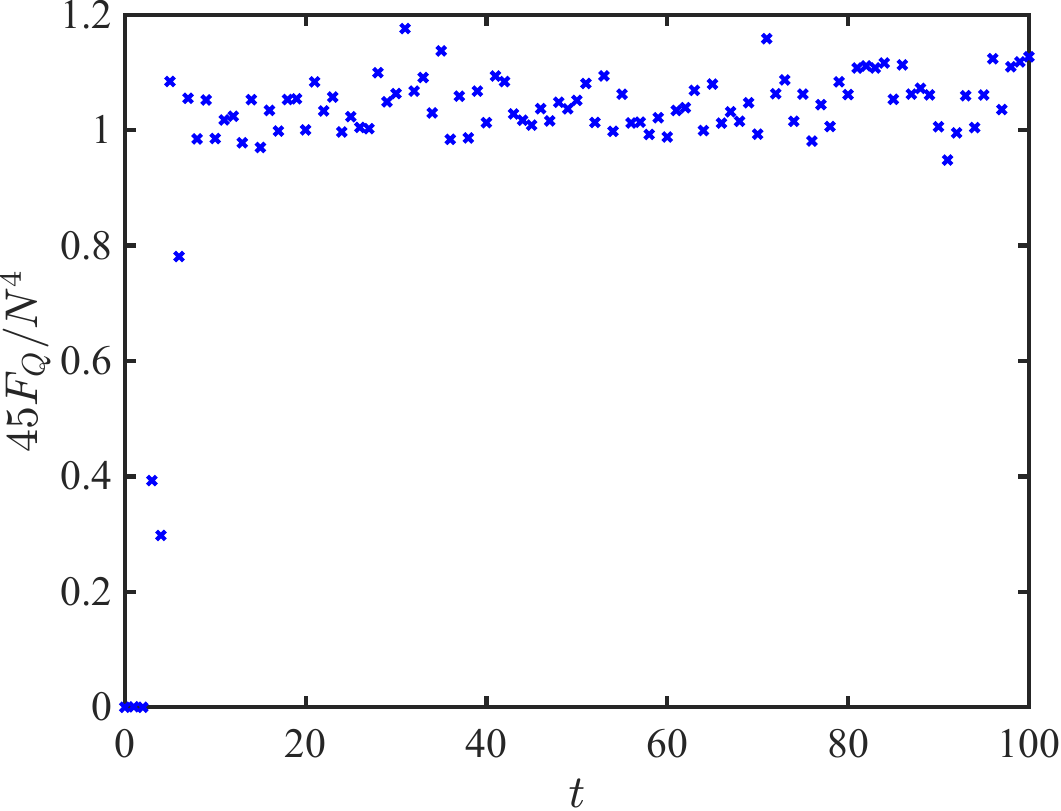}
\caption{
Evolution of QFI with respect to $J_z^2$ in the chaotic COE model with parameters $A=1.7$ and $C=10$.
}\label{figSM_COE}
\end{figure}

Furthermore, we can consider the case of taking $\hat O=\vect J_\alpha^2$.
Due to 
\begin{eqnarray}
    \tr_{\mathcal K_s}[\vect J_\alpha^4]=\sum_{m=-N/2}^{N/2} m^4=\frac{N(N+1)(N+2)(3N^2+6N-4)}{240},
\end{eqnarray}
then by Eq.~(\ref{SM-Main}), we have
\begin{eqnarray}\label{QFI-Jz^2}
    \bar{F}_{\rm chaos}[\vect J_\alpha^2]=\frac{N^4}{45}+\mathcal O(N^3), %
\end{eqnarray}
which is numerically verified in Fig.~\ref{figSM_COE}.

\section{ Calculation of Lyapunov Exponent (LE)}
The chaotic properties of quantum chaotic models are quantitatively assessed by their (largest) LE, denoted as $\lambda_{\rm LE}$, from their classical counterpart, which measures the rate of divergence between nearby orbits in classical dynamical systems. 
We take the COE-described kicked top model as an example to illustrate the numerical calculation of LE.
Its Hamiltonian is given by
\begin{eqnarray}\label{Floquet-COE}
&&U_{\rm COE}\!=\! \exp
\left(-i \frac{C}{N} \vect J_z^2\right) \exp\left(-i A \vect J_x\right),\\ \label{Floquet-CUE}
\end{eqnarray}
by defining classical variable $\vec X\equiv(X,Y,Z)=2(J_x,J_y,J_z)/N$, quantum dynamics $\vec J_{n+1}=U_{\rm COE}^\dag$ $\vec J_n U_{\rm COE} $ can be mapped onto classical dynamics in the limit $N\to\infty$, as follows
\begin{eqnarray}\label{Classical Dynamics of COE}
&&X_{n+1}=X_n\cos\Theta_n-
(\cos A Y_n-\sin A Z_n)
\sin\Theta_n,\n 
&&Y_{n+1}=X_n\sin\Theta_n+
(\cos A Y_n-\sin A Z_n)
\cos\Theta_n,\n 
&&Z_{n+1}=\cos A Z_n+\sin A Y_n,
\end{eqnarray}
where $\Theta_n=C(\cos A Z_n+\sin AY_n)$.
The LE can be calculated by using \cite{ParkerBook1990,ConstantoudisPRE1997,DArianoPRA1992}
\begin{eqnarray}\label{LE-num}
\lambda_{\rm LE}=\ln\left[\lim_{n\to\infty} (t_+(n))^{1/n}\right],
\end{eqnarray}
where $t_+(n)$ is the largest eigenvalue of $T=\prod_{p=1}^n \mathcal T(\vec X_p)$ and $\mathcal T(\vec X_n)=\partial \vec X_{n+1}/\partial  \vec X_n$.
In the main text, we will always consider the following coherent spin state as the initial state for numerical calculation,
\begin{eqnarray}\label{CSS}
\ket{\theta, \phi} &=& \cos^{N}\left(\frac{\theta}{2}\right)\exp\left[\tan\left(\frac{\theta}{2}\right)e^{i\phi}J^-\right]\ket{N/2,N/2} \n
&=& \bigotimes_{i=1}^N \left(\cos\frac{\theta}{2}\ket{\uparrow}_i+e^{i\phi}\sin\frac{\theta}{2}\ket{\downarrow}_i\right).
\end{eqnarray}
Therefore, the initial classical variables for coherent spin states are
\begin{eqnarray}
    X_0=\sin(\theta)\cos(\phi),  \ 
    Y_0=\sin(\theta)\sin(\phi),  \ 
    Z_0=\cos(\theta).
\end{eqnarray}
Equations~(\ref{Classical Dynamics of COE}) and (\ref{LE-num}) provide a numerical approach to calculate LE.

\section{An unstable point in the LMG model}
The Lipkin-Meshkov-Glick (LMG) model is given by
\begin{eqnarray}
H_{\rm LMG} = \Omega J_z -\frac{2\xi}{N}J_x^2.
\end{eqnarray}
To identify unstable points, we express the LMG Hamiltonian with respect to coherent spin states 
\begin{eqnarray}\label{Class-LMG}
h_{\rm LMG} &\equiv& \frac{\langle \theta,\phi|H_{\rm LMG}|\theta,\phi\rangle}{N/2} \n
&=& \Omega \cos\theta - 2\xi \sin^2\theta\cos^2\phi.
\end{eqnarray}
Canonical variables are defined as follows $Q = \sqrt{2(1+\cos\theta)}\cos\phi$ and $P = -\sqrt{2(1+\cos\theta)}$ $\sin\phi$. 
Consequently, Eq. (\ref{Class-LMG}) transforms into
\begin{eqnarray}
h_{\rm LMG} &=& \frac{\Omega}{2}(P^2+Q^2) - \Omega - \xi Q^2\left(1-\frac{P^2+Q^2}{4}\right).
\end{eqnarray}
Analysis of the classical Hamiltonian, Eq. (\ref{Class-LMG}), reveals that $P=Q=0$ denotes the unstable point when $\Omega(\Omega-2\xi)<0$ since $\partial_P h_{\rm LMG}(0,0)=\partial_Q h_{\rm LMG}(0,0)=0$ and $\partial_P^2 h_{\rm LMG}(0,0)$ $\partial_Q^2 h_{\rm LMG}(0,0)-[\partial_P\partial_Qh_{\rm LMG}(0,0)]^2=\Omega(\Omega-2\xi)<0$. 
Correspondingly, the positive Lyapunov Exponent (LE) for this unstable point, $P=Q=0$, is given by \cite{CameoPRE2020}
\begin{eqnarray}
\lambda_{\rm LE} = \sqrt{\Omega(2\xi-\Omega)}.
\end{eqnarray}

\begin{figure}[t]
\centering
\includegraphics[width=0.75\columnwidth]{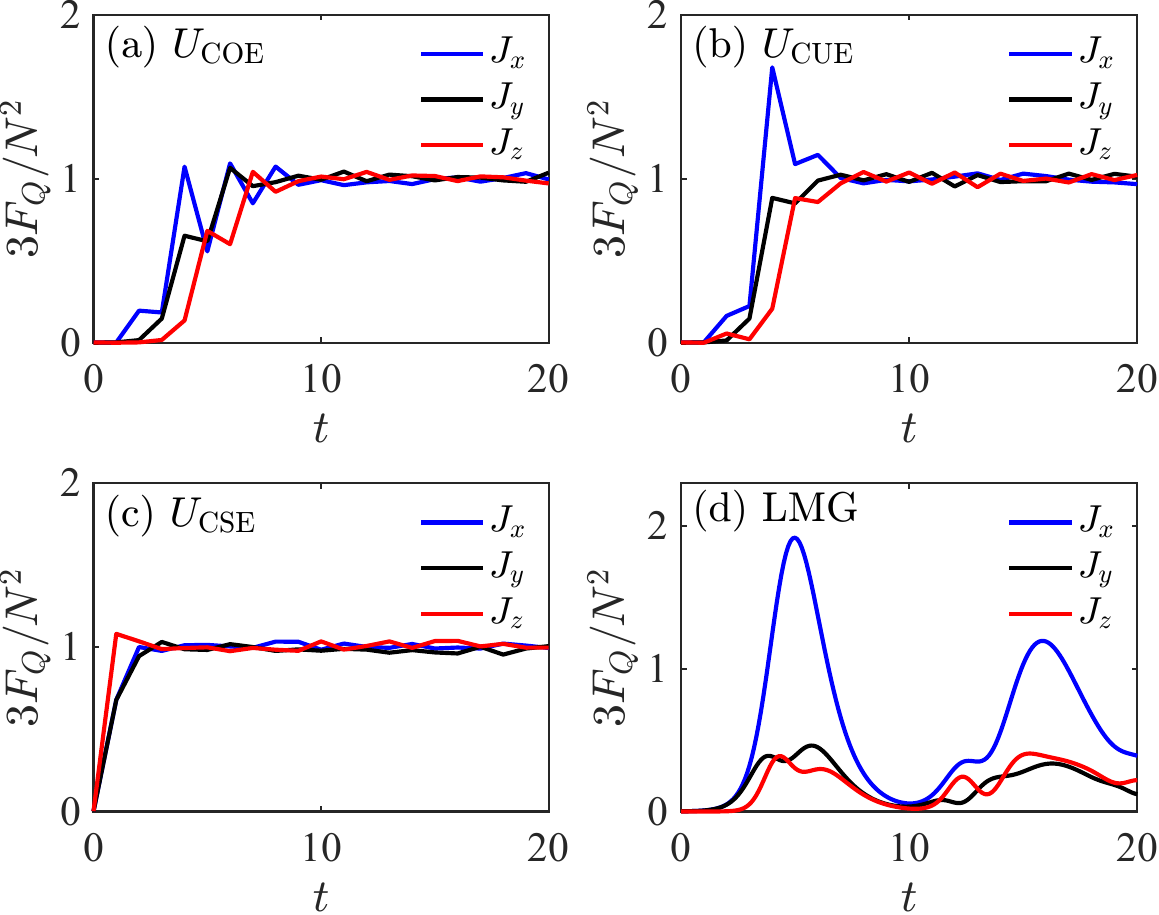}
\caption{
Evolution of QFI with respect to different spin operators $J_{x,y,z}$ in models of (a) $U_{\rm COE}$, (b) $U_{\rm CUE}$, (c) $U_{\rm CSE}$, and (d) LMG.
Chaotic conditions are used, $A=1.7, C=10$ for (a), $p = 1.7$, $\lambda=10$, $\lambda'=0.5$ for (b), and $\lambda_0= \lambda_1= 2.5$, $\lambda_2= 5$, and $\lambda_3= 7.5$ for (c).
The coherent spin state $\ket{\pi/2,-\pi/2}$ is used in panels (a-c).
In panel (d), we consider the unstable coherent spin state $\ket{\pi,\phi}$ as the initial state, and parameters of the LMG model are chosen as $\Omega=1$ and $\xi=-1$.
}\label{figSM}
\end{figure}

\section{QFI for different spin directions}

From Fig.~(\ref{figSM}), we see that in chaotic collective spin models, e.g., $U_{\rm COE}$, $U_{\rm CUE}$, and $U_{\rm CSE}$, defined in the main text, the short-time behavior of QFI is dependent on the choice of the direction of the spin operator $\vect J_\alpha$. However, in the long-time region, QFI in chaotic collective models behaves independently of $\alpha$ and tends towards a universal behavior, $F_Q \to N^2/3$.
For the LMG model, although at the unstable point, QFI displays exponential growth during the short time, there is no universal behavior of QFI even in the long-time region, and it highly depends on the choice of $\alpha$.

\end{appendix}


% Provide your bibliography here. You have two options:

%%% FIRST OPTION
% Write your entries here directly, following the example below, including:
% Author(s), Title, Journal Ref. with year in parentheses at the end, followed by the DOI number.

%\begin{thebibliography}{99}
%\bibitem{1931_Bethe_ZP_71} H. A. Bethe, {\it Zur Theorie der Metalle. i. Eigenwerte und Eigenfunktionen der linearen Atomkette}, Zeit. f{\"u}r Phys. {\bf 71}, 205 (1931), \doi{10.1007\%2FBF01341708}.
%\bibitem{arXiv:1108.2700} P. Ginsparg, {\it It was twenty years ago today... }, \url{http://arxiv.org/abs/1108.2700}.
%\end{thebibliography}

% Use your bibtex library, formatted by the SciPost style file.
%\bibliography{SciPost_Example_BiBTeX_File.bib}

\begin{thebibliography}{99}
%
\bibitem{GeorgescuRMP2014}
I. M. Georgescu, S. Ashhab, and F. Nori, Quantum simulation, Rev. Mod. Phys. \textbf{86}, 153 (2014), \doi{https://doi.org/10.1103/RevModPhys.86.153}
%
\bibitem{HorodeckiRMP2009}
R. Horodecki, P. Horodecki, M. Horodecki, and K. Horodecki, Quantum entanglement, Rev. Mod. Phys. \textbf{81}, 865 (2009), \doi{https://doi.org/10.1103/RevModPhys.81.865}
%
\bibitem{PezzeRMP2018}
L. Pezz\`e, A. Smerzi, M. K. Oberthaler, R. Schmied, and P. Treutlein, Quantum metrology with nonclassical states of atomic ensembles, Rev. Mod. Phys. \textbf{90}, 035005  (2018), \doi{https://doi.org/10.1103/RevModPhys.90.035005}
%
\bibitem{AmicoRMP2001}
L. Amico, R. Fazio, A. Osterloh, and V. Vedral,
Entanglement in many-body systems, Rev. Mod. Phys {\bf 80} 517 (2008), \doi{https://doi.org/10.1103/RevModPhys.80.517}
%
\bibitem{LaflorenciePhysRep2016}
N. Laflorencie, Quantum entanglement in condensed matter systems, Phys. Rep. \textbf{646}, 1 (2016), \doi{https://doi.org/10.1016/j.physrep.2016.06.008}

\bibitem{WittenRMP2018}
E. Witten, APS Medal for Exceptional Achievement in Research: Invited article on entanglement properties of quantum field theory, Rev. Mod. Phys. \textbf{90}, 045003  (2018), \doi{https://doi.org/10.1103/RevModPhys.90.045003}
%
\bibitem{RyuJHEP2006}
S. Ryu and T. Takayanagi, Aspects of holographic entanglement entropy, J. High Energy Phys. 08 (2006)
045, \doi{https://doi.org/10.1088/1126-6708/2006/08/045}
%
\bibitem{ArecchiPRA1972}
F. T. Arecchi, E. Courtens, R. Gilmore, and H. Thomas, Atomic Coherent States in Quantum Optics, Phys. Rev. A \textbf{6}, 2211 (1972), \doi{https://doi.org/10.1103/PhysRevA.6.2211}
%
%
\bibitem{KitagawaPRA1993}
M. Kitagawa and M. Ueda, Squeezed spin states,     Phys. Rev. A \textbf{47}, 5138  (1993), \doi{https://doi.org/10.1103/PhysRevA.47.5138}
%
\bibitem{DefenuPR2024}
N. Defenu, A. Lerose and S. Pappalardi,
Out-of-equilibrium dynamics of quantum many-body systems with long-range interactions, Phys. Rep. \textbf{1074}, 1 (2024), 
\doi{https://doi.org/10.1016/j.physrep.2024.04.005}
%
\bibitem{MicheliPRA2003}
A. Micheli, D. Jaksch, J. I. Cirac, and P. Zoller,
Many-particle entanglement in two-component Bose-Einstein condensates, Phys. Rev. A \textbf{67}, 013607 (2003), \doi{https://doi.org/10.1103/PhysRevA.67.013607}
%
\bibitem{MuesselPRA2015}
W. Muessel, H. Strobel, D. Linnemann, T. Zibold, B. Juliá-Díaz, and M. K. Oberthaler,
Twist-and-turn spin squeezing in Bose-Einstein condensates, Phys. Rev. A \textbf{92}, 023603 (2015), \doi{https://doi.org/10.1103/PhysRevA.92.023603}
%
\bibitem{SorelliPRA2019}
G. Sorelli, M. Gessner, A. Smerzi, and L. Pezz\`e, Fast and optimal generation of entanglement in bosonic Josephson junctions, Phys. Rev. A \textbf{99}, 022329 (2019), \doi{https://doi.org/10.1103/PhysRevA.99.022329}
%
\bibitem{HuangPRA2022}
J. Huang, H. Huo, M. Zhuang, and C. Lee,
Efficient generation of spin cat states with twist-and-turn dynamics via machine optimization, Phys. Rev. A \textbf{105}, 062456 (2022), \doi{https://doi.org/10.1103/PhysRevA.105.062456}
%
%\bibitem{MaldacenaJHEP2016}
%J. Maldacena, S. H. Shenker, and D. Stanford, A bound on chaos, \href{https://doi.org/10.1007/JHEP08(2016)106}{J. High Energy Phys. 08 (2016) 106.}
%
%\bibitem{LarkinSovJETP1969}
%A. Larkin and Y. N. Ovchinnikov, Quasiclassical Method in the Theory of Superconductivity, \href{http://www.jetp.ras.ru/cgi-bin/dn/e_028_06_1200.pdf}{Sov. Phys. JETP \textbf{28}, 1200 (1969).}
%
%
\bibitem{PezzePRL2009}
L. Pezz\`e and A. Smerzi, Entanglement, Nonlinear Dynamics, and the Heisenberg Limit, Phys. Rev. Lett. \textbf{102}, 100401  (2009), \doi{https://doi.org/10.1103/PhysRevLett.102.100401}
%
\bibitem{HyllusPRA2012}
P. Hyllus, W. Laskowski, R. Krischek, C. Schwemmer, W. Wieczorek, H. Weinfurter, L. Pezz\`e, and A. Smerzi, Fisher information and multiparticle entanglement,  Phys. Rev. A \textbf{85}, 022321 (2012), \doi{https://doi.org/10.1103/PhysRevA.85.022321  }
%
\bibitem{TothPRA2012}
G. T\'oth, Multipartite entanglement and high-precision metrology,    Phys. Rev. A \textbf{85}, 022322 (2012), \doi{https://doi.org/10.1103/PhysRevA.85.022322}
%
%\bibitem{ImaiARXIV2024}
%S. Imai, A. Smerzi, and L. Pezz\`e, Metrological usefulness of entanglement and nonlinear Hamiltonians, \href{https://doi.org/10.48550/arXiv.2405.15703}{arXiv:2405.15703.}
%
\bibitem{StrobelSCIENCE2014}
H. Strobel, W. Muessel, D. Linnemann, T. Zibold, D. B. Hume, L. Pezzè, A. Smerzi, and M. K. Oberthaler,
Fisher information and entanglement of non-Gaussian spin states, Science {\bf 345}, 424 (2014), \doi{https://doi.org/10.1126/science.1250147}
%
\bibitem{BohnetSCIENCE2016}
J. G. Bohnet, B. C. Sawyer, J. W. Britton, M. L. Wall, A. M. Rey, M. Foss-Feig, and J. J. Bollinger,
Quantum spin dynamics and entanglement generation with hundreds of trapped ions, Science 352, 1297 (2016), \doi{https://doi.org/10.1126/science.aad9958}
%
\bibitem{GarttnerPRL2018}
M. G\"arttner, P. Hauke, and A. M. Rey, Relating Out-of-Time-Order Correlations to Entanglement via Multiple-Quantum Coherences, Phys. Rev. Lett. \textbf{120}, 040402 (2018), \doi{https://doi.org/10.1103/PhysRevLett.120.040402}
%
\bibitem{PappalardiPRA2018}
S. Pappalardi, A. Russomanno, B. Žunkovič, F. Iemini, A. Silva, and R. Fazio,
Scrambling and entanglement spreading in long-range spin chains, Phys. Rev. B \textbf{98}, 134303 (2018), \doi{https://doi.org/10.1103/PhysRevB.98.134303}
%
\bibitem{LerosePRA2020}
A. Lerose and S. Pappalardi, Bridging entanglement dynamics and chaos in semiclassical systems,  Phys. Rev. A \textbf{102}, 032404 (2020), \doi{https://doi.org/10.1103/PhysRevA.102.032404}
%
\bibitem{WimbergerBook2014}
S. Wimberger, {\it Nonlinear dynamics and quantum chaos} (Springer, Berlin, 2014), \doi{https://doi.org/10.1007/978-3-319-06343-0}
%
\bibitem{TrainPRL2021} M. C. Tran, A. Y. Guo, C. L. Baldwin, A. Ehrenberg, A. V. Gorshkov, and A. Lucas, 
Lieb-Robinson light cone for power-law interactions, Phys. Rev. Lett. \textbf{127}, 160401 (2021), \doi{https://doi.org/10.1103/PhysRevLett.127.160401}
%
\bibitem{GuoPRA2020} A.Y.Guo, M.C.Tran, A.M.Childs, A.V.Gorshkov, and Z.-X. Gong, 
Signaling and scrambling with strongly long-range interactions, Phys. Rev. A \textbf{102}, 010401 (2020), \doi{https://doi.org/10.1103/PhysRevA.102.010401}
%
\bibitem{ShiPRL2024}
H.-L. Shi, X.-W. Guan, and J. Yang, Universal Shot-Noise Limit for Quantum Metrology with Local Hamiltonians, Phys. Rev. Lett.
\textbf{132}, 100803 (2024), \doi{https://doi.org/10.1103/PhysRevLett.132.100803}
%
\bibitem{ChuPRL2023}
Y. Chu, X. Li, and J. Cai, Strong Quantum Metrological Limit from Many-Body Physics,     Phys. Rev. Lett. \textbf{130}, 170801 (2023), \doi{https://doi.org/10.1103/PhysRevLett.130.170801}
%
\bibitem{ChuArXiv2024}
Y. Chu, X. Li, J. Cai, Quantum delocalization on correlation landscape: The key to exponentially fast multipartite entanglement generation, Phys. Rev. Lett. \textbf{133}, 110201 (2024), \doi{https://doi.org/10.1103/PhysRevLett.133.110201}
%
\bibitem{DysonMathPhys1962-2}
F. J. Dyson, Statistical theory of the energy levels of
complex systems. II, J. Math. Phys. (N.Y.) \textbf{3}, 157 (1962), \doi{https://doi.org/10.1063/1.1703774}
%
\bibitem{DysonMathPhys1962-3}
F. J. Dyson, Statistical theory of the energy levels of
complex systems. III, J. Math. Phys. (N.Y.) \textbf{3}, 166 (1962), \doi{https://doi.org/10.1063/1.1703775}
%
\bibitem{DysonMathPhys1962-1}
F. J. Dyson, Statistical theory of the energy levels of
complex systems. I, J. Math. Phys. (N.Y.) \textbf{3}, 140 (1962), \doi{https://doi.org/10.1063/1.1703773}
%
\bibitem{GoldsteinEPJH2010}
S. Goldstein,  J. L. Lebowitz, R. Tumulka, and N. Zanghi, Long-time behavior of macroscopic quantum systems,  Eur.
Phys. J. H \textbf{35}, 173 (2010), \doi{https://doi.org/10.1140/epjh/e2010-00007-7}
%
\bibitem{PolkovnikovRMP2011}
A. Polkovnikov, K. Sengupta, A. Silva, and M. Vengalattore, Colloquium: Nonequilibrium dynamics of closed interacting quantum systems, Rev. Mod. Phys. \textbf{83}, 863 (2011), \doi{https://doi.org/10.1103/RevModPhys.83.863}
%
\bibitem{NandyARXIV}
P. Nandy, A. S. Matsoukas-Roubeas, P. Martínez-Azcona, A. Dymarsky, and A. del Campo,
Quantum Dynamics in Krylov Space: Methods and Applications, arXiv:2405.09628, \doi{https://doi.org/10.48550/arXiv.2405.09628}
%
%
\bibitem{FN0}
Collective spin models in this Letter refer to the quantum systems whose Hamiltonians are described solely by collective spin operators $J_{\alpha}$.
%
%
\bibitem{ChangPRA2021}
S.-C. Li, L. Pezz\`e, and A. Smerzi,
Multiparticle entanglement dynamics of quantum chaos in a Bose-Einstein condensate,
Phys. Rev. A \textbf{103}, 052417 (2021), \doi{https://doi.org/10.1103/PhysRevA.103.052417}
%
\bibitem{note_thermalization} Equation~(\ref{Main}) can be rewritten as $\bar F_{\rm chaos}\!\simeq\! 4(\Delta \hat O)_{\mathbb I_\mathcal K/K}$ with $\mathbb I_\mathcal K/K$ being the infinite temperature thermal state in the Krylov space, which is related to quantum thermalization, see M. Brenes, S. Pappalardi, J. Goold, and A. Silva, Multipartite Entanglement Structure in the Eigenstate Thermalization Hypothesis, Phys. Rev. Lett. \textbf{124}, 040605  (2020), \doi{https://doi.org/10.1103/PhysRevLett.124.040605}
%
\bibitem{OszmaniecPRX2016}
M. Oszmaniec, R. Augusiak, C. Gogolin, J. Ko\l ody\'nski, A. Ac\'in, and M. Lewenstein, Random Bosonic States for Robust Quantum Metrology, Phys. Rev. X \textbf{6}, 041044 (2016),  \doi{https://doi.org/10.1103/PhysRevX.6.041044}
%
\bibitem{FN-PRX}
The result $\bar{F}_{\rm rand}[\hat O]\!=\!4\tr_{\mathcal K}[\hat O^2]/(K\!+\!1)$ is found in Eq.~(C12) of Ref.~\cite{OszmaniecPRX2016}.
There, the condition $\tr_{\mathcal K}[O]\!=\!0$ is assumed.
Nevertheless, the result of Ref.~\cite{OszmaniecPRX2016} can be directly extended to the case of $\tr_{\mathcal K}[\hat O]\!\neq\!0$ by replace the origin $\hat O$ with $\hat O\!-\!\tr_{\mathcal K}[\hat O]/K$ and lead to $\bar{F}_{\rm rand}[\hat O]\!=\!4\tr_{\mathcal K}[\hat O^2]/(K\!+\!1)-4(\tr_{\mathcal K}[\hat O])^2/K(K+1)$, which  agrees with our Eq.~(\ref{Main}) in the limit $K\gg 1$.
%
\bibitem{RobertsJHEP2017}
D. A. Roberts and B. Yoshida, Chaos and complexity by design, J. High Energ. Phys. \textbf{2017}, 121 (2017), \doi{https://doi.org/10.1007/JHEP04(2017)121}
%
\bibitem{EmersonScience2003}
J. Emerson, Y. S. Weinstein, M. Saraceno, S. Lloyd, and D. G.
Cory, Pseudo-random unitary operators for quantum information processing, Science \textbf{302}, 2098 (2003), \doi{https://doi.org/10.1126/science.1090790}
%
\bibitem{NakataPRX2017}
Y. Nakata, C. Hirche, M. Koashi, and A. Winter,
Efficient Quantum Pseudorandomness with Nearly Time-Independent Hamiltonian Dynamics, Phys. Rev. X {\bf 7}, 021006 (2017), \doi{https://doi.org/10.1103/PhysRevX.7.021006}
%
\bibitem{ChoiNATURE2023}
J. Choi, et al., 
Preparing random states and benchmarking with many-body quantum chaos, Nature {\bf 613}, 468 (2023), \doi{https://doi.org/10.1038/s41586-022-05442-1}
%
\bibitem{ChaudhuryNATURE2009}
S. Chaudhury, A. Smith, B. Anderson, S. Ghose, and P. S. Jessen,
Quantum signatures of chaos in a kicked top, Nature {\bf 461}, 768 (2009), 
\doi{https://doi.org/10.1038/nature08396}
%
\bibitem{NeillNP2016}
C. Neill, et al., Ergodic dynamics and thermalization in an isolated quantum system, Nat. Phys. \textbf{12}, 1037–1041 (2016), \doi{https://doi.org/10.1038/nphys3830}
%
\bibitem{KrithikaPRE2019}
V. R. Krithika, V. S. Anjusha, U. T. Bhosale, and T. S. Mahesh, NMR studies of quantum chaos in a two-qubit kicked top, Phys. Rev. E \textbf{99}, 032219 (2019), \doi{https://doi.org/10.1103/PhysRevE.99.032219}
%
\bibitem{ChalopinNC2018}
T. Chalopin, C. Bouazza, A. Evrard, V. Makhalov, D. Dreon, J. Dalibard, L. A. Sidorenkov, and S. Nascimbene,  Quantum-enhanced sensing using non-classical spin states of a highly magnetic atom, Nat. Commun. \textbf{9}, 4955 (2018), \doi{https://doi.org/10.1038/s41467-018-07433-1}
%
\bibitem{MourikPRE2018} 
V. Mourik, S. Asaad, H. Firgau, J. J. Pla, C. Holmes, G. J. Milburn, J. C. McCallum, and A. Morello, Exploring quantum chaos with a single nuclear spin, Phys. Rev. E \textbf{98}, 042206 (2018), \doi{https://doi.org/10.1103/PhysRevE.98.042206}
%
\bibitem{BrittonNature2012}
J. W. Britton, B. C. Sawyer, A. C. Keith, C.-C. J. Wang, J. K. Freericks, H. Uys, M. J. Biercuk, and J. J. Bollinger, Engineered two-dimensional Ising interactions in a trapped-ion quantum simulator with hundreds of spins, Nature \textbf{484}, 489–492 (2012), \doi{https://doi.org/10.1038/nature10981}
%
\bibitem{GarttnerNP2017}
M. G\"{a}rttner,  J. G. Bohnet, A. Safavi-Naini, M. L. Wall, J. J. Bollinger, and A. M. Rey, Measuring out-of-time-order correlations and multiple quantum spectra in a trapped-ion quantum magnet, Nat. Phys. \textbf{13}, 781–786 (2017), \doi{https://doi.org/10.1038/nphys4119}
%
\bibitem{BornetNATURE2023}
G. Bornet, et al., 
Scalable spin squeezing in a dipolar Rydberg atom array, Nature {\bf 621}, 728 (2023), \doi{https://doi.org/10.1038/s41586-023-06414-9}

\bibitem{MarkPRL2023}
D. K. Mark, J. Choi, A. L. Shaw, M. Endres, and S. Choi,
Benchmarking Quantum Simulators Using Ergodic Quantum Dynamics, Phys. Rev. Lett. {\bf 131}, 110601 (2023), \doi{https://doi.org/10.1103/PhysRevLett.131.110601}
%
\bibitem{LantanoARXIV}
See also a recent proposal using sequential measurements of a central spin coupled with a spin ensemble and reaching an average QFI $\sim N^2/3$ ($\sim N$) for symmetric (non-symmetric) coupling:
T. B. Lantan\~o, D. Yang, K. M. R. Audenaert, S. F. Huelga, and M. B. Plenio, 
Unlocking Heisenberg Sensitivity with Sequential Weak Measurement Preparation, 
arXiv:2403.05954, \doi{https://doi.org/10.48550/arXiv.2403.05954}
%
\bibitem{MacriPRA2016}
T. Macr\`i, A. Smerzi, and L. Pezz\`e, Loschmidt echo for quantum metrology,    Phys. Rev. A \textbf{94}, 010102(R) (2016), \doi{https://doi.org/10.1103/PhysRevA.94.010102}
%
\bibitem{FN1}
OTOCs capture the complexity arising from the non-commutativity of operators $W_{\alpha}(\theta,t)$ and $\rho(0)$, linked to the LE.
%
Explicitly, for $t\!\lesssim\! t^*$,  both OTOC and QFI typically grow exponentially at the rate of the LE from their classical analogues.
%
Moreover, within this time regime, the non-commutativity depends on the direction $\alpha$ of the spin operator $J_{\alpha}$, and thus QFI also varies with the choice of $\alpha$, reflecting the fact that the evolved state is not fully randomized, see Fig.~\ref{figWigner}(a-b).
%
%
\bibitem{ZurekSCIENCE2001}
W. H. Zurek, Sub-Planck structure in phase space and its relevance for quantum decoherence, Nature  \textbf{412}, 712–717 (2001), \doi{https://doi.org/10.1038/35089017}
%
\bibitem{ToscanoPRA2006}
F. Toscano, D. A. R. Dalvit, L. Davidovich, and W. H. Zurek,
Sub-Planck phase-space structures and Heisenberg-limited measurements
Phys. Rev. A 73, 023803 (2006), \doi{https://doi.org/10.1103/PhysRevA.73.023803}
%%
%
%\bibitem{supp} See SM for the detailed derivation of Eq.~(\ref{Main}) and some deduced results by considering different Krylov spaces.
%
Detailed discussions on the calculation of LEs for kicked-top models, the LMG model, and the dependence of QFI on spin directions can also be found in SM.
%
\bibitem{TothJPA2014}
G. T\'oth and I. Apellaniz, Quantum metrology from a quantum information science perspective, J. Phys. A \textbf{47}, 424006 (2014), \doi{https://doi.org/10.1088/1751-8113/47/42/424006}
%
%
\bibitem{BraunsteinPRL1994}
S. L. Braunstein and C. M. Caves, Statistical distance and the geometry of quantum states, Phys. Rev. Lett. \textbf{72}, 3439 (1994), \doi{https://doi.org/10.1103/PhysRevLett.72.3439}
%

\bibitem{HongPRA2015}
Y. Hong, S. Luo, and H. Song, Detecting $k$-nonseparability via quantum Fisher information, Phys. Rev. A \textbf{91}, 042313  (2015), \doi{https://doi.org/10.1103/PhysRevA.91.042313}


%
\bibitem{RenPRL2021}
Z. Ren, W. Li, A. Smerzi, and M. Gessner, Metrological Detection of Multipartite Entanglement from Young Diagrams, Phys. Rev. Lett. \textbf{126}, 080502  (2021), \doi{https://doi.org/10.1103/PhysRevLett.126.080502}
%

\bibitem{HaakeBook2010}
F. Haake, Quantum Signatures of Chaos (Springer Science
\& Business Media, New York, 2010), Vol. 54, \doi{https://doi.org/10.1007/978-3-642-05428-0}
%
\bibitem{MehtaBook2004}
M. L. Mehta, Random Matrices, 3rd ed. (Academic Press,
Cambridge, MA, 2004), \doi{https://doi.org/10.1016/C2009-0-22297-5}
%

%
%\bibitem{BaumgratzPRL2014}
%T. Baumgratz, M. Cramer, and M. B. Plenio, Quantifying Coherence, \href{https://doi.org/10.1103/PhysRevLett.113.140401}{Phys. Rev. Lett. \textbf{113}, 140401 (2014).}
%
%
\bibitem{UllahPR1965}
N. Ullah and C. E. Porter, Expectation-Value Distributions, Phys. Rev. \textbf{137}, B1394 (1965), \doi{https://doi.org/10.1103/PhysRev.137.B1394}
%
\bibitem{DAlessioAP2016}
L. D'Alessio, Y. Kafri, A. Polkovnikov, and M. Rigol,
From quantum chaos and eigenstate thermalization to statistical mechanics and thermodynamics,
Advances in Physics \textbf{65}, 239 (2016), \doi{https://doi.org/10.1080/00018732.2016.1198134}
%
\bibitem{FN2}
The time-reversal symmetry $T^2\!=\!1$~($T^2\!=\!-1$) of models~(\ref{Floquet-COE}) [~(\ref{Floquet-CSE}) for odd $N$] implies that orthogonal (symplectic) transformations as its canonical transformations. 
Hence, these models belong to the COE (CSE) classification.
%
In the absence of additional conserved quantities besides parity symmetry and $\vect{J}^2$, unitary transformations serve as canonical transformations for model~(\ref{Floquet-CUE}), placing it under the CUE classification.
%
\bibitem{FN3}
For the CSE model~(\ref{Floquet-CSE}), the LE has not been found due to the lack of classical analogues, as far as we know.
%
%
\bibitem{NohPRE2021}
J. D. Noh, Operator growth in the transverse-field Ising spin chain with integrability-breaking longitudinal field, Phys. Rev. E \textbf{104}, 034112 (2021), \doi{https://doi.org/10.1103/PhysRevE.104.034112}
%
\bibitem{KarthikPRA2007}
J. Karthik, A. Sharma, and A. Lakshminarayan, Entanglement, avoided crossings, and quantum chaos in an Ising model with a tilted magnetic field, Phys. Rev. A \textbf{75}, 022304 (2007), \doi{https://doi.org/10.1103/PhysRevA.75.022304}
%
\bibitem{XiaARXIV2024}
W. Xia, J. Zou, and X. Li, Complexity enriched dynamical phases for fermions on graphs, arXiv:2404.08055, \doi{https://doi.org/10.48550/arXiv.2404.08055}
%
\bibitem{CameoPRE2020}
S. Pilatowsky-Cameo, J. Ch\'avez-Carlos, M. A. Bastarrachea-Magnani, P. Str\'ansk\'y, S. Lerma-Hern\'andez, L. F. Santos, and J. G. Hirsch, Positive quantum Lyapunov exponents in experimental systems with a regular classical limit, Phys. Rev. E \textbf{101}, 010202 (2020), \doi{https://doi.org/10.1103/PhysRevE.101.010202}
%
\bibitem{HummelPRL2019}
Q. Hummel, B. Geiger, J. D. Urbina, and K. Richter, Reversible Quantum Information Spreading in Many-Body Systems near Criticality, Phys. Rev. Lett. \textbf{123}, 160401 (2019),  \doi{https://doi.org/10.1103/PhysRevLett.123.160401}
%

\bibitem{LipkinNuclearP1965}
H. J. Lipkin, N. Meshkov, and A. Glick, Validity of many-body approximation methods for a solvable model: (I). Exact solutions and perturbation theory, Nuclear Physics \textbf{62}, 188 (1965), \doi{https://doi.org/10.1016/0029-5582(65)90862-X}
%
\bibitem{MeshkovNuclearP1965}
N. Meshkov, A. Glick, and H. Lipkin, Validity of many-body approximation methods for a solvable model: (II). Linearization procedures, Nuclear Physics \textbf{62}, 199 (1965),  \doi{https://doi.org/10.1016/0029-5582(65)90863-1}
%
\bibitem{GlickNuclearP1965}
A. Glick, H. Lipkin, and N. Meshkov, Validity of many-body approximation methods for a solvable model: (III). Diagram summations, Nuclear Physics \textbf{62}, 211 (1965), \doi{https://doi.org/10.1016/0029-5582(65)90864-3}
%
\bibitem{OrtizNPB2005}
G. Ortiz, R. Somma, J. Dukelsky, and S. Rombouts, Exactly-Solvable Models Derived from a Generalized Gaudin Algebra, Nucl.Phys.  B \textbf{707}, 421 (2005), \doi{https://doi.org/10.1016/j.nuclphysb.2004.11.008}
%
\bibitem{FerreiraPRA2013}
A. G. Araujo-Ferreira, R. Auccaise, R. S. Sarthour, I. S. Oliveira, T. J. Bonagamba, and I. Roditi, Classical bifurcation in a quadrupolar NMR system, Phys. Rev. A \textbf{87}, 053605 (2013), \doi{https://doi.org/10.1103/PhysRevA.87.053605}
%
\bibitem{ZirnbauerJMP1996}
M. R. Zirnbauer, Riemannian symmetric superspaces and their origin in random‐matrix theory, J. Math. Phys. \textbf{37}, 4986 (1996), \doi{https://doi.org/10.1063/1.531675}
%
\bibitem{AltlandPRB1997}
A. Altland and M. R. Zirnbauer, Nonstandard symmetry classes in mesoscopic normal-superconducting hybrid structures, Phys. Rev. B \textbf{55}, 1142 (1997), \doi{https://doi.org/10.1103/PhysRevB.55.1142}
%
\bibitem{AltlandPReport2002}
A. Altland, B. Simons, and M. Zirnbauer, Theories of low-energy quasi-particle states in disordered d-wave superconductors
Author links open overlay panel, Physics Reports \textbf{359}, 283 (2002), \doi{https://doi.org/10.1016/S0370-1573(01)00065-5}
%
\bibitem{ZirnbauerArXiv2004}
M. R. Zirnbauer, Symmetry classes in random matrix theory, arXiv: math-ph/0404058 (2004), \doi{https://doi.org/10.48550/arXiv.math-ph/0404058}
%
\bibitem{HeinznerCMP2005}
P. Heinzner, A. Huckleberry, and M. R. Zirnbauer, Symmetry classes of disordered fermions, Commun.Math.Phys.  \textbf{257}, 725 (2005), \doi{https://doi.org/10.1007/s00220-005-1330-9} 
%
\bibitem{KawabataPRXQ2023}
K. Kawabata, A. Kulkarni, J. Li, T. Numasawa, and S. Ryu, Symmetry of Open Quantum Systems: Classification of Dissipative Quantum Chaos, PRX Quantum \textbf{4}, 030328 (2023), \doi{https://doi.org/10.1103/PRXQuantum.4.030328}
%
\bibitem{AugusiakPRA2012}
We emphasize that the QFI detect metrologically-useful entanglement~\cite{PezzeRMP2018, PezzePRL2009}. It has been shown that entangled symmetric states of $N$ qubits are genuine multipartite entangled, see R. Augusiak, J. Tura, J. Samsonowicz, and M. Lewenstein, Entangled symmetric states of $N$ qubits with all positive partial transpositions, Phys. Rev. A \textbf{86}, 042316 (2012), \doi{https://doi.org/10.1103/PhysRevA.86.042316}


%%%%%%%%%%%%%%


\bibitem{ParkerBook1990}
T. S. Parker and L. O. Chua, Practical Numerical Algorithms for Chaotic Systems (Springer-Verlag, New York,
1990), \doi{https://doi.org/10.1007/978-1-4612-3486-9}
%
\bibitem{ConstantoudisPRE1997}
V. Constantoudis and N. Theodorakopoulos, Lyapunov exponent, stretching numbers, and islands of stability of the kicked top, Phys. Rev. E \textbf{56}, 5189 (1997), \doi{https://doi.org/10.1103/PhysRevE.56.5189}
%
\bibitem{DArianoPRA1992}
G. M. D'Ariano, L. R. Evangelista, and M. Saraceno, Classical and quantum structures in the kicked-top model, Phys. Rev. A \textbf{45}, 3646 (1992), \doi{https://doi.org/10.1103/PhysRevA.45.3646}
%



\end{thebibliography}

%%%%%%%%%% END TODO: BIBLIOGRAPHY

\end{document}